\PassOptionsToPackage{table}{xcolor}
\documentclass[acmlarge]{acmart}

\usepackage{booktabs} % For formal tables
\usepackage{wrapfig}
\usepackage{tikz} % For tikz drawings
\usepackage{multirow}

\usepackage[ruled]{algorithm2e} % For algorithms

\SetAlFnt{\small}
\SetAlCapFnt{\small}
\SetAlCapNameFnt{\small}
\SetAlCapHSkip{0pt}
\IncMargin{-\parindent}
\usepackage{subcaption}

\setcopyright{rightsretained} 
\acmJournal{IMWUT}
\acmVolume{3}
\acmNumber{3}
\acmArticle{83}
\acmYear{2019}
\acmMonth{9}

\acmPrice{}
\acmDOI{10.1145/3351241}

\newcommand{\quotepar}[2]{\begin{quote}{\textit{#1}} #2\end{quote}}

\begin{document}
% Title portion
\title{Blocks: Collaborative and Persistent Augmented Reality Experiences}
% \titlenote{}

\author{Anhong Guo}
\affiliation{
\institution{Human-Computer Interaction Institute, Carnegie Mellon University}
\city{Pittsburgh}
\state{PA}
\postcode{15213}
}
\email{anhongg@cs.cmu.edu}

\author{Ilter Canberk}
\affiliation{
\institution{Snap Inc.}
\city{Venice}
\state{CA}
\postcode{90291}
}
\email{ilter@snap.com}

\author{Hannah Murphy}
\affiliation{
\institution{Wellesley College}
\city{Wellesley}
\state{MA}
\postcode{02481}
}
\email{hannah.murphy@wellesley.edu}

\author{Andr\'es Monroy-Hern\'andez}
\affiliation{
\institution{Snap Inc.}
\city{Seattle}
\state{WA}
\postcode{98121}
}
\email{amh@snap.com}

\author{Rajan Vaish}
\affiliation{
\institution{Snap Inc.}
\city{Santa Monica}
\state{CA}
\postcode{90405}
}
\email{rvaish@snap.com}

\begin{abstract}
We introduce {\em Blocks}, a mobile application that enables people to co-create AR structures that persist in the physical environment. Using Blocks, end users can collaborate synchronously or asynchronously, whether they are colocated or remote. Additionally, the AR structures can be tied to a physical location or can be accessed from anywhere. We evaluated how people used Blocks through a series of lab and field deployment studies with over 160 participants, and explored the interplay between two collaborative dimensions: space and time. We found that participants preferred creating structures synchronously with colocated collaborators. Additionally, they were most active when they created structures that were not restricted by time or place. Unlike most of today's AR experiences, which focus on content consumption, this work outlines new design opportunities for persistent and collaborative AR experiences that empower anyone to collaborate and create AR content.
\end{abstract}




\begin{teaserfigure}
\centering
\includegraphics[width=0.85\textwidth]{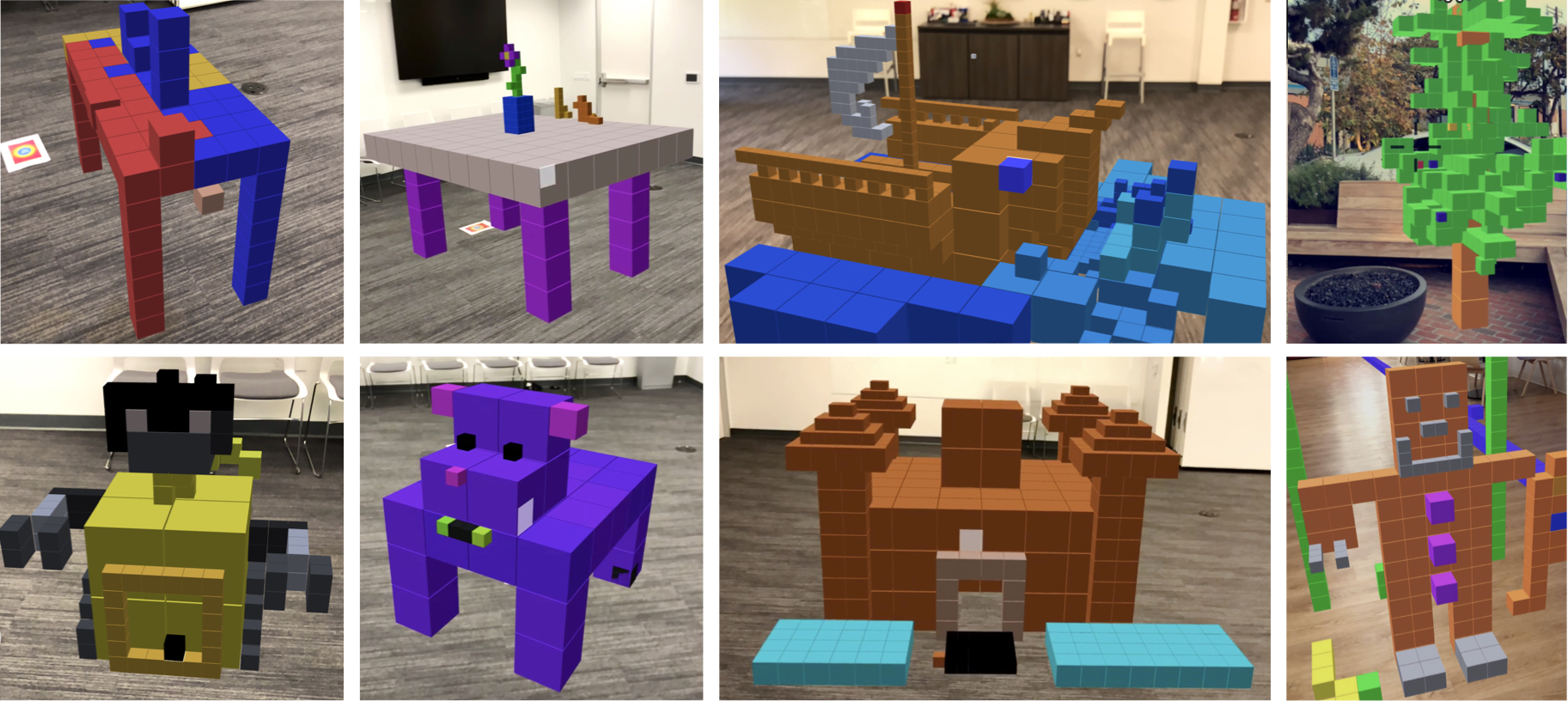}
\caption{AR structures showing the diversity of creations people built collaboratively using Blocks. Here we highlight a few structures, including two tables, a sailboat, a tree, a robot, a dog, a castle and a gingerbread man.}
\label{fig:teaser}
\end{teaserfigure}

\maketitle

\renewcommand{\shortauthors}{Guo et al.}

\section{Introduction}\label{sec:intro}
Augmented reality (AR) experiences are now accessible to millions of people worldwide through a variety of mobile applications such as Snapchat \cite{Snapchat} and Pok\'emon Go \cite{PokemonGo}. However, today's AR landscape is at the same stage as the Internet was about two decades ago, when only experts could create the content that end users consumed. For example, only expert 3D animators can create virtual characters in popular AR games like Pok\'emon Go. The advent of the so-called Web 2.0 brought a range of platforms that enabled end users to create, collaborate, and publish content. Web 2.0 shifted the paradigm from consumption to creation \cite{benkler2006wealth, jenkins2006convergence}. Similarly, we envision a future of collaborative creation in AR, where everyone can contribute to the digital content, making information more accessible and embedded in the physical world.

In this paper, we introduce Blocks, a mobile application that enables anyone to co-create persistent AR structures using cubes or `blocks.' We developed the feature set of Blocks based on the two dimensions in the collaborative AR systems literature: space and time \cite{brockmann2013framework}. As depicted in Table~\ref{tab:DesignSpace}, Blocks enables collaboration by allowing multiple people to build structures together, synchronously or asynchronously. The structures persist, i.e., they do not disappear from session to session. These structures are either tied to the physical location where the user created them (location-dependent) or exist as ``free-floating'' structures in a shared AR world (location-independent). Lastly, this persistence facilitates colocated and remote collaboration. To our knowledge, Blocks is the first AR application that works across all of the design dimensions of space and time in a single tool, making it possible to understand their effect on end-user experience and participation.

We evaluated Blocks through a 24-person lab study, and a 68-person field study on a company campus, focusing on examining people's experiences and participation across the design space in Table~\ref{tab:DesignSpace}. The lab study consisted of twelve participant pairs. Each pair used Blocks during three 15-minute sessions. In some sessions participants were colocated, while in others they were remote. The field study lasted for three days and consisted of recruiting participants as they passed by common areas on a company campus. Participants used Blocks for many short sessions (5 minutes on average) throughout the deployment, during which we experimented with location-dependent and independent scenarios.

\begin{wraptable}{r}{0.55\textwidth}
\vspace*{-3mm}
\caption{We examined the interplay between two design dimensions in collaborative AR: time and space. \textit{Time} in terms of synchronous and asynchronous collaboration. \textit{Space} in terms of location of the AR content and the creators (i.e., whether they are colocated or remote).}
\label{tab:DesignSpace}
\begin{center}
\def\checkmark{\tikz\fill[scale=0.5](0,.25) -- (.25,0) -- (0.65,.7) -- (.25,.15) -- cycle;}
\renewcommand\arraystretch{1.2}
\setlength{\arrayrulewidth}{0.8pt}
\small
\vspace{-0.5pc}
\begin{tabular}{|c|c|c|c|c|}\cline{3-4}
\multicolumn{2}{c|}{}&\multicolumn{2}{c|}{\textbf{Collaboration}}&\multicolumn{1}{c}{}\\\cline{3-4}
\multicolumn{2}{c|}{}& \multicolumn{1}{c|}{Sync}& Async &\multicolumn{1}{c}{}\\\hline
\multirow{2}{*}{\begin{tabular}{@{}c@{}}\textbf{Location of} \\[-4pt]\textbf{people}\end{tabular}}& Colocated & \checkmark & N/A & \multirow{2}{*}{Lab Study} \\\cline{2-4}
& Remote & \checkmark & \checkmark &\\\hline
\multirow{2}{*}{\begin{tabular}{@{}c@{}}\textbf{Location of} \\[-4pt]\textbf{AR structures}\end{tabular}}& Dependent & \checkmark & \checkmark & \multirow{2}{*}{Field Study}\\\cline{2-4}
& Independent & \checkmark & \checkmark & \\\hline
\end{tabular}
\end{center}
\end{wraptable}

Participants created a wide variety of structures, from tables to castles, from robots to sailboats, using more than 12,000 blocks in the lab study and close to 7,000 blocks in the field study (see Figure~\ref{fig:teaser}). In terms of quality experience, participants in the lab study reported enjoying synchronous colocated collaboration the most. In terms of high activity, participants engaged with the app during more sessions throughout the day in the location-independent scenario of the field study. Overall, we found it promising that many participants in the lab study willingly spent more time working on their creations than required, and many in the field study repeatedly engaged with Blocks. Finally, we conducted a five-day extended deployment with 70 participants in a naturalistic setting, and our findings were consistent with the other two studies.

The main contributions of this paper are: (i) the design and implementation of Blocks, a mobile application that enables collaborative and persistent mobile augmented reality experiences across multiple design dimensions, (ii) empirical results from lab and field deployment studies that advance our understanding of how varying design dimensions affect end-user experience and participation, and (iii) recommendations for designing engaging experiences aligned with a vision of a world where anyone can co-create persistent AR content.

\section{Related Work}\label{sec:relatedwork}
Current research and commercial AR systems primarily focus on user experiences that are consumption-centric, non-collaborative, or non-persistent. In this paper, we explore novel design configurations around AR experiences where collaborative creations persist in the physical space. Blocks works across all of these design dimensions in a single tool, making it easier to compare and study different styles and modes of collaborative authoring. 

\subsection{AR Content Creation}
The availability of AR software development toolkits, such as Apple's ARKit \cite{ARKit} and Google's ARCore \cite{ARCore}, have contributed to an explosion of mobile AR applications. However, most of those applications are centered around consuming AR content created by experts, confining end-users' role to lightweight interactions with AR objects. For example, commercial applications like Pok\'emon Go \cite{PokemonGo} or research projects like Brick \cite{brick}, focus mainly on enabling users to interact with, rather than create AR content.

There are research \cite{wang2013diy,ha2010artalet,roussos1999learning,Lee:2004,Lee:2005} and commercial systems (e.g., wiARframe \cite{wiARframe}, ZapWorks \cite{ZapWorks}, and Maquette \cite{Maquette}) for creating AR/VR content using direct manipulation and WYSIWYG interfaces. However, these systems were developed with designers and expert creators in mind. 

Inspired by how prior work on construction kits for children has democratized the creation of 2D digital content \cite{Resnick:2005}, we too focus on broadening AR creation to non-experts. In this work we focus on exploring ways to ``lower the floor'' of AR creation by making the experience of building AR structures as approachable as building them with LEGO bricks. Furthermore, we also aim to enable people to create as complex and diverse AR structures as they want: from a simple tower to a complex sailboat. We envision a future where people's creativity can be expressed and overlaid onto the physical world through AR.

\subsection{Collaboration in AR}
There are several systems, such as Arrow \cite{arrow} and Experiment with Google's Garden Friends \cite{GardenFriends}, that enable users to create AR content. However, these systems focus on single users, rather than collaborative interactions. Even beyond collaborative AR creation, researchers have identified the need for more formal studies on collaborative AR systems in general \cite{zhou2008trends,kim2018revisiting}. For instance, a meta analysis of the AR literature found that only 10 of the 161 AR studies they surveyed focused on collaboration \cite{Billinghurst:2015}. Another meta analysis identified a set of design dimensions for collaborative AR applications: space, time, mobility, virtual content, user roles, and visualization hardware \cite{brockmann2013framework}. In this work, we focus on two of those dimensions that are particularly relevant for collaborative creation \cite{remixedreality}: {\em time} and {\em space} (see Table~\ref{tab:DesignSpace}). 

Studies on collaboration modalities \cite{pan2018empowerment} examined synchronous collaboration between two users in different mixed reality settings and found that AR-to-AR ones were associated with greater collaboration, embodiment, presence, and co-presence. 
Similarly, researchers \cite{kiyokawa2002communication, henrysson2005face} have compared colocated and remote collaboration in AR. They found that physical world visibility had significant impact on communication and awareness. However, these systems did not fully explore the design landscape we articulated in Table~\ref{tab:DesignSpace}. As a result, we cannot compare how different design choices in that table affect user experiences. Our contribution is to examine these choices through a single application. 

\subsection{Role of Location in AR}
The widely adopted game Minecraft \cite{Minecraft} enables people to collaboratively create virtual structures using 3D blocks. More recently, the Minecraft team showed experimental AR versions of the Minecraft application for head-mounted displays and mobile devices. We drew inspiration from this and focused on taking the experience out of the boundaries of a desktop/mobile game and into the world by enabling people to have their structures persist at specific locations. We explored this in a more open-ended creation environment, outside the constraints of a video game \cite{muller2015statistical}, allowing us to evaluate a wide range of design choices. 

Prior research has investigated the opportunities of having AR content linked to specific locations \cite{paucher2010location,Reitmayr:2003,starner1997augmented} in the areas of navigation support \cite{Reitmayr:2003}, museum tours \cite{damala2008bridging}, and interactive games such as Pok\'emon Go \cite{PokemonGo}. Similarly, location-independent AR content is popular in applications such as Snapchat's AR lenses \cite{Snapchat} and IKEA Place \cite{IKEAPlace}. In our work, we focus on the role of location when it comes to enabling users to create AR structures that are dependent and independent of a specific location. 

\section{Blocks}\label{sec:blocks}
Blocks is an iOS mobile application that enables people to co-create AR structures that can persist in the physical environment (see Figure~\ref{fig:System}). We designed the app to use blocks or voxels to create AR structures because they are like 3D pixels for the AR world. There are at least two widely successful systems that use cubes as the basic building unit: LEGO \cite{lego} and Minecraft \cite{Minecraft}. Inspired by these systems, we bring what people love about them to AR to support end-user creation. Furthermore, we focus on mobile AR rather than other platforms because smartphones are ubiquitous, allowing the deployment of Blocks in the wild. In this section, we describe a set of features that enable users to onboard, collaborate, and create persistent AR structures. 

\subsection{Core Features}
We applied basic design principles from the literature and existing systems (e.g., construction kits for children \cite{Resnick:2005}, collaborative writing tools \cite{mabrito2006study, Olson:2017}, and block-based creative environments \cite{lego,Minecraft}) to design the tools and interactions for supporting creation and collaboration in AR. We conducted formative studies with 12 participants during seven sessions. We asked the participants to think aloud while creating structures with Blocks. After each session, we improved and refined Blocks based on participant's feedback.

\begin{figure}[t!]
 \centering
 \hfill
 \begin{subfigure}[b]{0.63\linewidth}
 \centering
 \includegraphics[width=\textwidth]{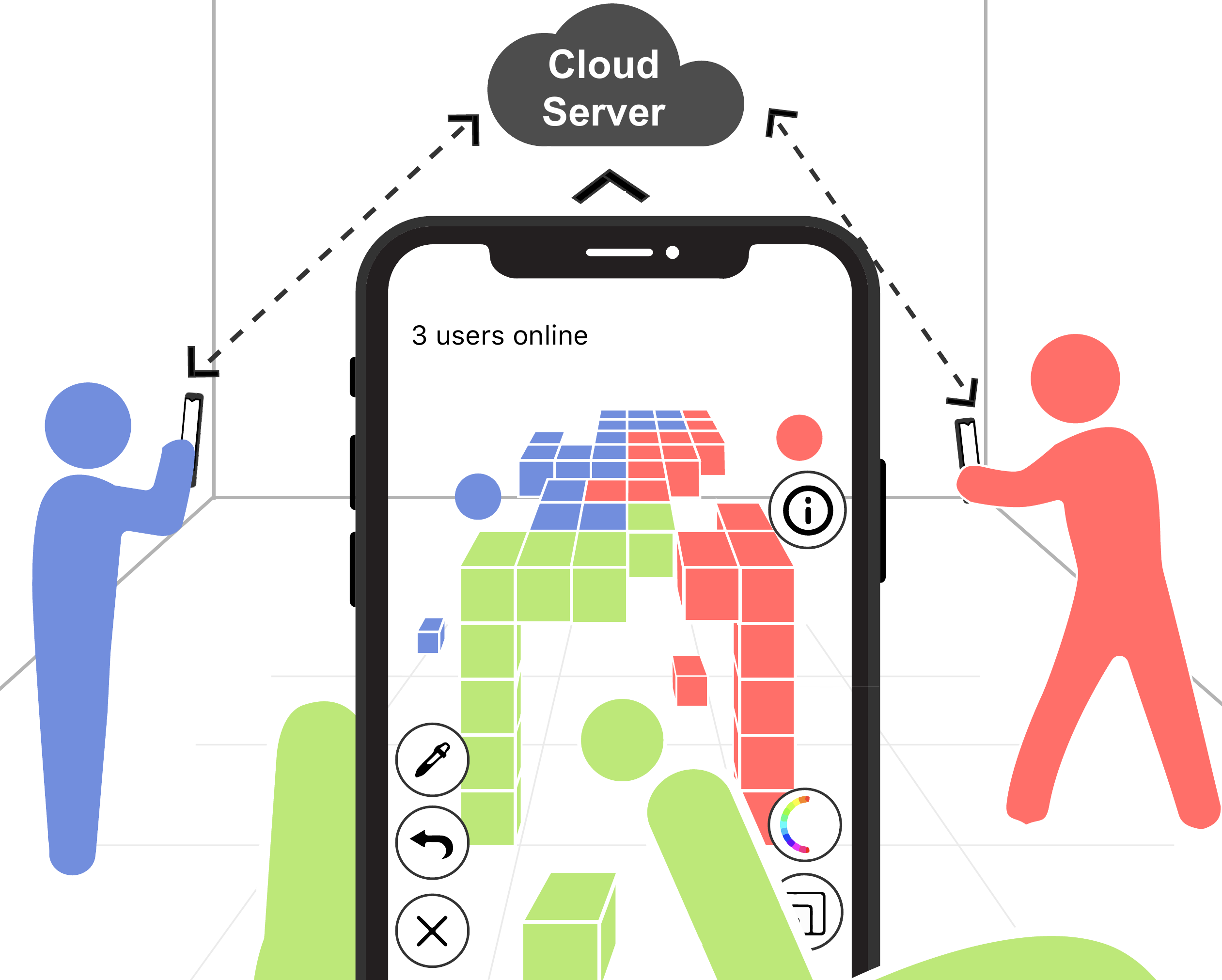}
 \end{subfigure}
 \hfill
 \begin{subfigure}[b]{0.33\linewidth}
 \centering
 \includegraphics[width=\textwidth]{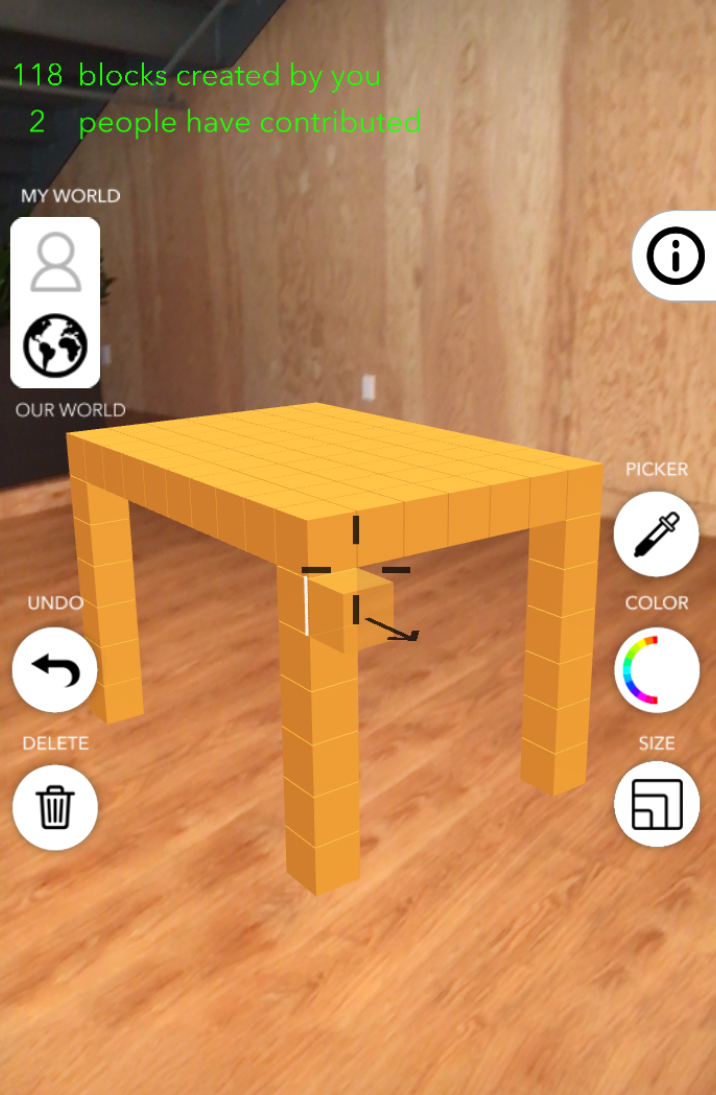}
 \end{subfigure}
 \hfill
 \vspace{-0.5pc}
 \caption{An illustration of three people co-creating a table in a colocated setup using Blocks (left), and a screenshot showing the Blocks user interface (right). The interface provides functionalities to add a block, pick a color and size, delete or undo blocks, and see collaborators' cursors. Note that Blocks can also be used for remote collaboration when people are not in the same location.}
\label{fig:System}
\vspace{-1pc}
\end{figure}

Through the formative studies, we settled on the following UI features (Figure~\ref{fig:System}): 
\vspace{-0.5pc}
\begin{itemize}
 \item Add a block (tap on the screen): Users can add a block on real-world surfaces or next to existing blocks.
 \item Add multiple blocks (press and hold on the screen): Users can create columns and rows of blocks.
 \item Display usage status (Info panel): Users can see the number of blocks added and users online.
 \item Undo a block (press the Undo button): Users can undo the last block added by themselves.
 \item Delete a block (press the Delete button): Users can aim at an existing block added by anyone and delete it. 
 \item Instructional hints (press the `i' button): Users can learn about Blocks' features through hints.
 \item Pick color (use the Color Picker): Users can aim at an existing block and pick its color.
 \item Change color (use the Color Panel): Users can change color from a continuous color range panel.
 \item Select size (use the Size Picker): Users can select the size of the blocks from three options.
 \item Show cursors (in the AR scene): Users can see each others' cursors. The cursor shows the exact size and color of the block they are placing.
\end{itemize}

Additionally, Blocks plays a sound effect when adding or deleting a block; highlights recent changes with a lighter color for a short period of time; preserves blocks' properties (e.g., position, size, color, and ownership) over time; and supports location-dependence with shared image markers.

\subsection{Creation}
Blocks embraces Resnick's `low floor' principle \cite{Resnick:2005} by making it easy to get started: simply aiming with the cursor and tapping on the screen places a virtual block in the environment captured by the camera. Similarly, Blocks embodies the `high ceiling' principle by not having predefined limits on what people can create. Like LEGO or Minecraft, the ceiling is set by people's own imagination and time constraints. Lastly, Blocks realizes the `wide walls' principle by enabling people to create a wide diversity of structures: from colorful flowers to robots, to dragons, and more. Blocks enables people to think of their world as their canvas, and offers the AR-equivalent of a pixel.

Using `world tracking' in ARKit, Blocks recognizes real-world surfaces in the camera's field of view so users can place blocks on top of them. Users can also add new blocks next to existing ones through ray casting \cite{roth1982ray}. Furthermore, users can add columns or rows of blocks by pressing-and-holding on the screen continuously. The direction of the column or row is determined by the ray-casted surface when the press-and-hold gesture happens, which can be either vertical or horizontal. Users can delete a block by aiming the cursor on an existing block and tapping on the delete button. Users can also use the undo button to remove the last block they added. Additionally, Blocks plays sound effects when adding or deleting a block to enhance the tapping experience.

Blocks provides three options for the length of the edge of the cubes: roughly 2, 4, and 8 centimeters. We chose these three sizes based on the sizes of everyday physical objects: an ice cube, a Rubik's Cube, and a gift box. During our formative studies, we found that these sizes were sufficient for users to create structures at different scales: small enough to create decorative details and large enough to quickly create large structures. Blocks also offers a continuous range of colors for users to choose from. Also informed by our formative studies, Blocks randomly selects a default color to encourage colorful creations. 

\begin{figure}[t!]
 \centering
 \hfill
 \begin{subfigure}[b]{0.33\linewidth}
 \centering
 \includegraphics[width=\textwidth]{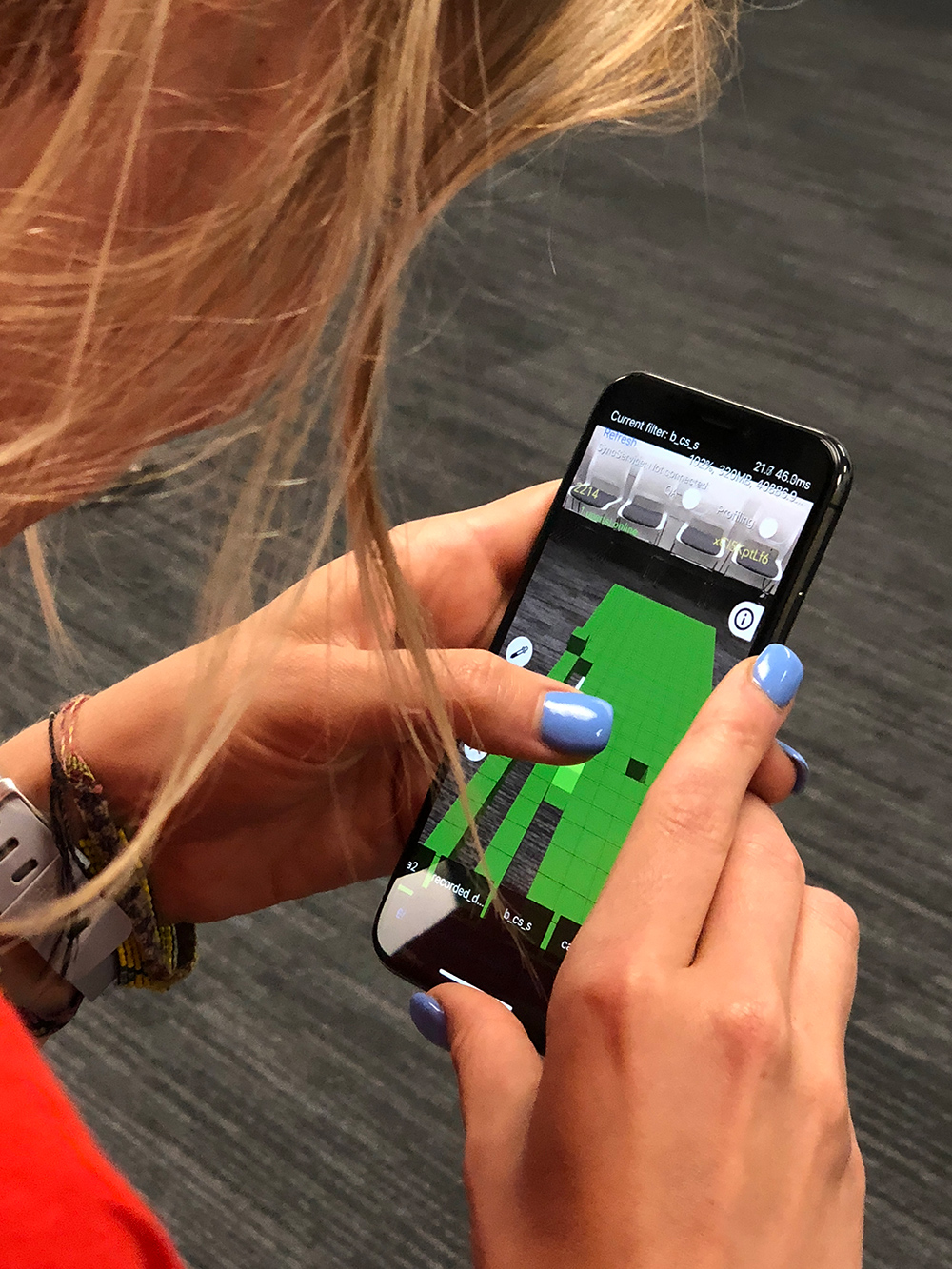}
 \end{subfigure}
 \hfill
 \begin{subfigure}[b]{0.645\linewidth}
 \centering
 \includegraphics[width=\textwidth]{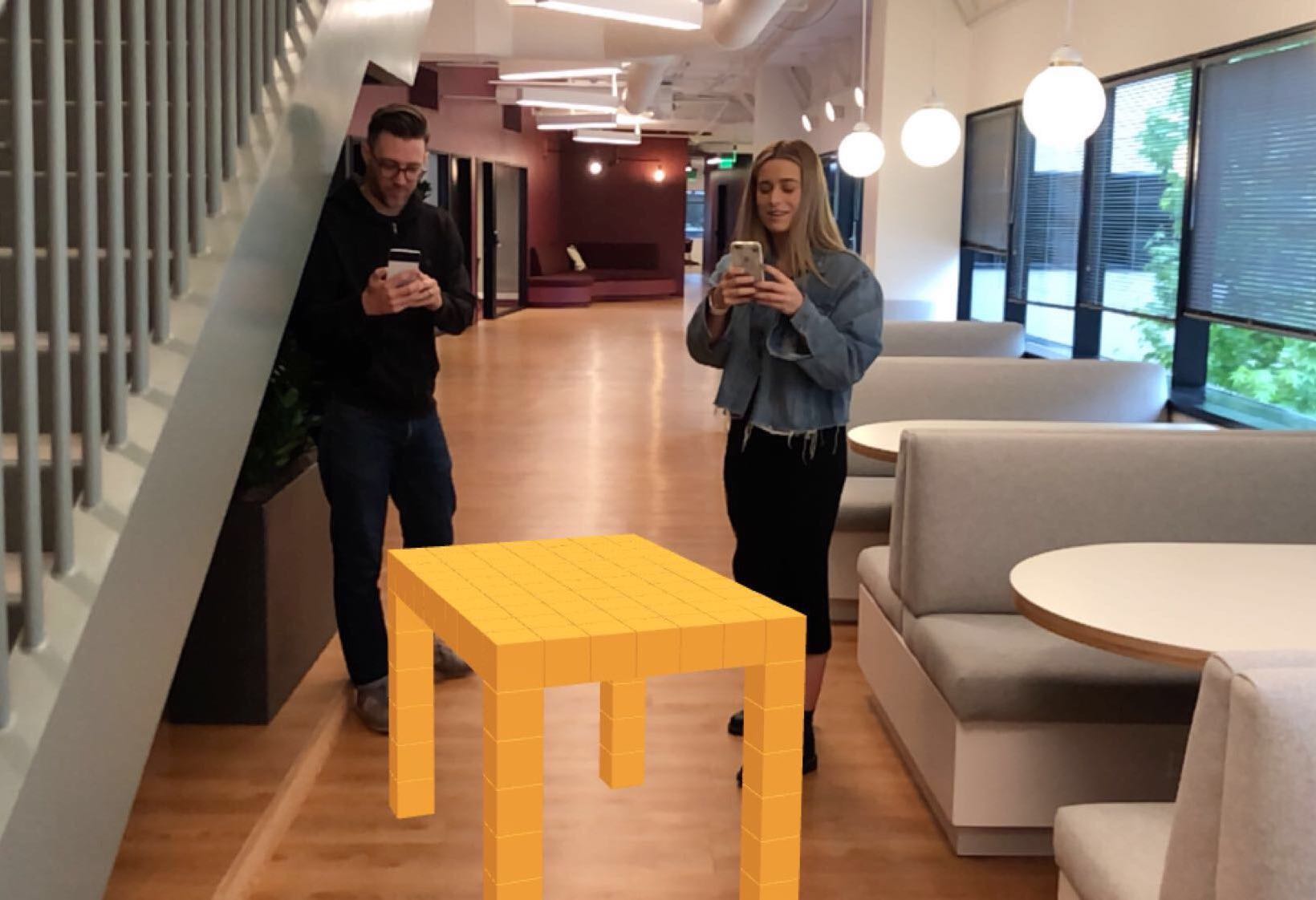}
 \end{subfigure}
 \hfill
 \vspace{-0.5pc}
 \caption{A user using Blocks to create a layer of green lawn (left), and two users collaboratively creating a table (right).}
\label{fig:UsersUsing}
\vspace{-1pc}
\end{figure}

\subsection{Collaboration}
Blocks enables collaboration through a back-end component that maintains the states of the different users, maps, and blocks. We implemented Blocks' back end using Node.js, deployed it on Google Cloud infrastructure, and used Firebase's Cloud Firestore as our database, since it supports real-time event streaming. Blocks enables persistence of the virtual blocks by keeping track of their position, size, color, ownership (in the form of an anonymous, randomly generated ID) and other properties. The Blocks client app updates the AR `mesh' to reflect the latest changes.

Inspired by collaborative editing tools such as Google Docs \cite{GoogleDocs}, Blocks synchronizes and shows online users' cursors. Users can see others' cursors with the exact size and color they are using at that moment. Blocks shows the number of potential collaborators that are online.

In our formative studies we noticed that users often talked to each other when they added blocks. They did this to coordinate and get feedback because they realized the other users might not have noticed their changes. Blocks highlights the recently added blocks with a lighter color for a short period of time (1.5s). We also noticed users trying to mimic their partner's color as they built their structures, only later realizing the color was slightly different. In its final design, Blocks integrates a color picker to support such collaborative behavior. Of course, users can also use the color picker to select a color they had used before.

\subsection{Onboarding Experience}
Blocks integrates several techniques to introduce new users to the app (onboarding). When the app launches, Blocks shows a screen with instructions and hints for different features, allowing the users to try them out. Blocks also applies the parallel existence of a personal AR world (``MyWorld'') and a shared AR world (``OurWorld''). The distinction of the two worlds allows new users to first try out the features in their personal world as a sandbox, reducing the risk of unintentionally ruining or sabotaging the shared AR world.

Users differ in their willingness to build or alter other peoples' creations. Inspired by how Sketch-a-bit \cite{tuite2012emergent} gave users the choice to either get a new drawing or start from a blank one, Blocks onboards users with instructions like ``build on others or start your own.'' Furthermore, inspired by approaches found in the literature on Legitimate Peripheral Participation \cite{lave1991situated}, Blocks gets users comfortable with editing in OurWorld by, for example, showing an incomplete structure, e.g., a table, with a few missing and redundant blocks in the hopes that it can scaffold their participation in the community.

\subsection{Location-Dependence of AR Structures}
Prior approaches for linking AR structures to physical locations include the use of GPS \cite{PokemonGo}, fiducial markers \cite{cai2014case,billinghurst2000mixing}, reference images \cite{Guo:2016,Guo:2017}, and magnetic fields \cite{Rajagopal:2018}. Blocks uses images (e.g., a poster on the wall) as markers to maintain the orientation, rotation, and scale of the virtual structures relative to the physical world across sessions. When Blocks detects an image marker, it fetches the structures registered to that marker's coordinate system. Subsequently, each time a marker is re-identified, the virtual structures are re-calibrated to correct any errors in the world tracking. Using this approach, if the marker moves, the structures moves with it too. Note that image-marker tracking is used for multi-user experiences and works across sessions. On the other hand, world tracking is used to locate real-world surfaces and maintain the image-marker tracking coordinates when the image is not visible in the camera view.

\section{Lab Study}\label{sec:labstudy}
The goal of our lab study was two-fold: (i) to evaluate the effectiveness of the tools and interactions designed to support end-user creation and collaboration in AR, and (ii) to investigate how creation and collaboration differ when people are colocated vs. remote, as indicated in Table~\ref{tab:DesignSpace}.

\subsection{Participants}
We recruited 24 participants (13 male, 11 female) from the technology company where the authors work. We focused on recruiting participants from other departments and with job titles that included engineer, designer, product manager, technician, and financial analyst. Users could enroll to participate by themselves or in pairs with a friend or colleague of their choice. Participants skewed young, with six of them aged 18 to 24 years old, 17 aged 25 to 34 years old, and one in the range of 35-44 years old. The majority of the participants had a bachelor's degree or had completed some college (83\%). Most participants had some experience with AR and VR: seventeen participants self-reported having limited AR/VR experience and seven reported having extensive experience and had designed AR-based experiences as part of their jobs.

\begin{figure}[t!]
 \centering
 \vspace{-1pc}
 \includegraphics[width=0.8\linewidth]{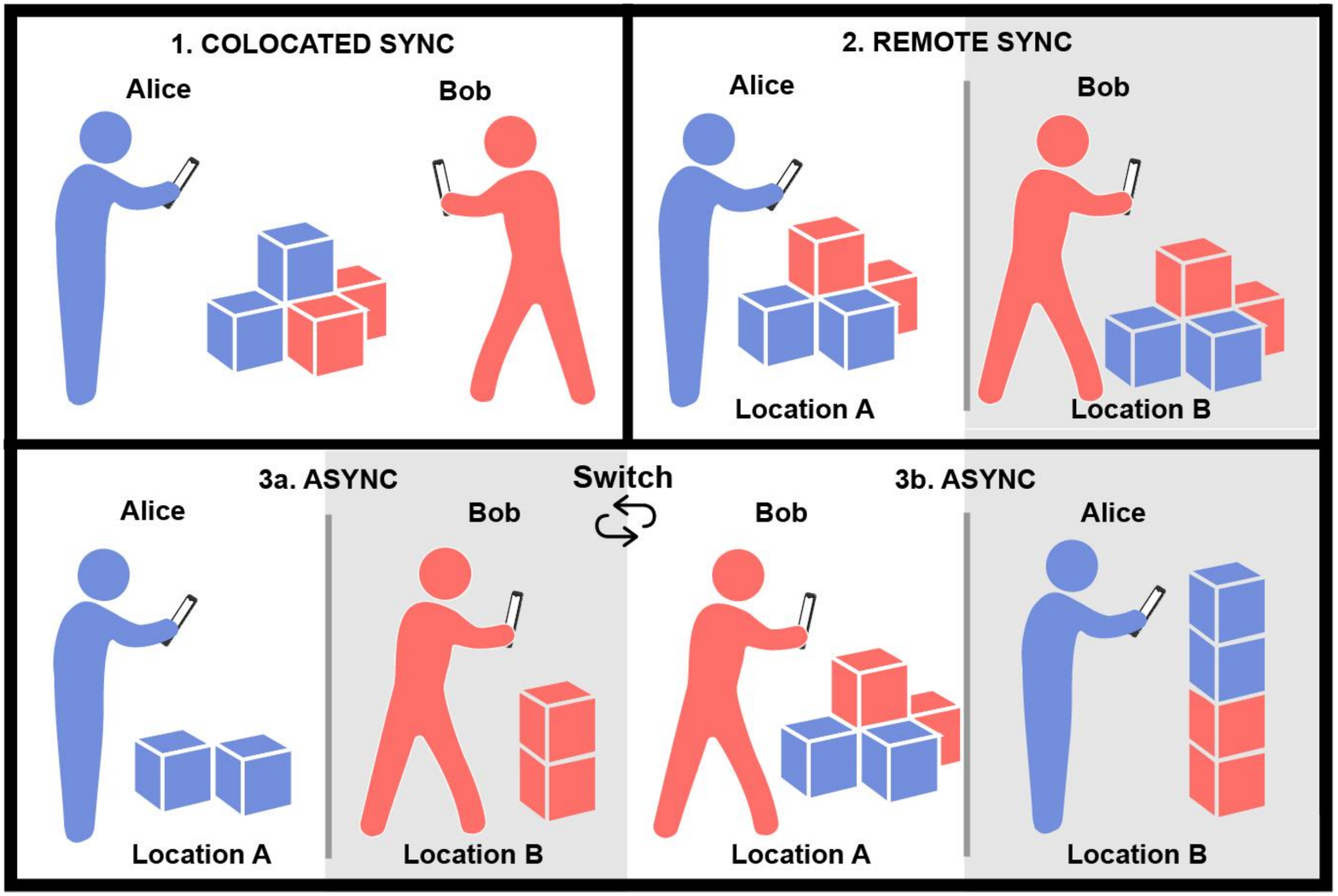}
 \vspace{-1pc}
 \caption{Experimental setup for the lab study. In the \textit{colocated-sync} condition (1), a pair of participants were physically in the same environment, collaborating on the same structure. In the \textit{remote-sync} condition (2), a pair of participants were in two separated environments, while collaborating on the same structure. In the \textit{async} condition (3), each participant first individually created part of their own structure, then switched to build on their partner's structure.}
 \vspace{-1pc}
\label{fig:LabSetup}
\end{figure}

\subsection{Study Method}
Our study used a within-subject design in which two participants created block-based structures with three different modes of collaboration and location of people in a counterbalanced order. As indicated in the Lab Study part of Table~\ref{tab:DesignSpace}, the three conditions were: (i) {\em colocated-sync}, where participants collaborated on the same structure at the same time and in the same place, so they could communicate both verbally and visually (Figure~\ref{fig:LabSetup}.1), (ii) {\em remote-sync}, where participants collaborated on the same structure at the same time, but in two separate environments, so they could only communicate verbally (Figure~\ref{fig:LabSetup}.2), and (iii) {\em async}, where participants first individually created part of their own structure in each environment then switched to build on their partner's structure in the place the structure was created (Figure~\ref{fig:LabSetup}.3). As a result, participants in two different places created two structures. In this condition, they could not communicate at the time of creation but were allowed to exchange thoughts during the transition. Note that for all three conditions, participants collaboratively created structures for the same amount of time.

For collaborative and persistent AR experiences, these conditions represent all the possible scenarios across time and space. For instance, in other collaborative systems, people can collaborate at different times (async, e.g., writing, construction), together at different places (remote-sync, e.g., video-conferencing), or together at the same place (colocated-sync, e.g., pair programming). Our study covers these collaboration modes through a single application, enabling us to understand the outcomes, processes, and their implications for building compelling collaborative AR experiences.

\subsection{Procedure and Setup}
Following the informed consent and a brief introduction of the study, participants answered a pre-study survey about demographics, past experience, and their in-the-moment happiness, creativity, and closeness level with the other study participant, using the Inclusion of Other in the Self scale (IOS) \cite{aron1992inclusion}. We gave participants two iPhone X devices running iOS 11.4.1 with Blocks installed and we showed a few examples of virtual sculptures created in the formative studies. We then instructed participants to familiarize themselves with Blocks in a free-form exploratory session of 5-10 minutes.

Next, we asked each pair of participants to collaboratively create block-based structures in three 15-minute sessions with counterbalanced order of collaboration modes. We instructed them to build a table, then a robot, and finally a structure of their choice. We fixed the order of the goals to increase the difficulty and openness as time progressed. In each session, we encouraged participants to plan their structure in any format they preferred. They could stop at any time, or could continue creating structures after the 15-minute had passed. We told participants that their structure would be public and persistent at that location, so other Blocks users who came later could see and build on it. 

The lab study sessions were conducted in a large empty room that could seat 80 people with a separator in the middle, so that they could communicate verbally when in two separate environments. After each session, participants were asked to complete a short survey about their experience. The surveys featured open-ended responses and 7-point Likert scale ratings including ``I had fun'', ``I feel engaged'', and ``I'm satisfied with what I/we built.'' Participants also left notes via a Google Form after each session for future creators who might build on their creation. At the end of the study, participants were asked to answer a post-study survey, specifically focused on their overall experience, ratings of the creation and collaboration features of Blocks, and comparison of the three conditions. For all Likert scale questions, participants rated along a scale of 1 to 7, where 1 was extremely negative and 7 was extremely positive.

The studies were video recorded from both sides of the room. Server logs of blocks creation and deletion were recorded and used for further analysis. In total, each study took approximately 90 minutes. Because we measured users' engagement and spontaneous play, we did not provide monetary incentives. At the end of the study, pictures and videos of the participants' creations were shared with them as souvenirs to show their friends and colleagues.

\section{Lab Study Results}\label{sec:labstudyresults}
To study the effect of people's location on creation and collaboration, we investigated the group outcome, the collaborative process, and participants' subjective feedback. To evaluate how well Blocks performed as a creation and collaboration tool, we investigated the subjective ratings and open-ended responses for the features of Blocks.

We analyzed quantitative results for our within-subject studies using parametric one-way repeated measures ANOVA and non-parametric Friedman test. We used an alpha level of 0.05. Additionally, our post-hoc tests used paired t-tests and Wilcoxon signed-rank tests with Holm-Bonferroni method for correcting multiple comparisons (the reported \textit{p} values were adjusted). We treated Likert scales with 7 points as approximating equal intervals and thus analyzed them using ANOVAs or T-tests.\footnote{Note that our results were consistent when tested with non-parametric equivalents (Friedman and Wilcoxon signed-rank tests) to our tests.} For open-ended responses, we conducted a thematic analysis \cite{guest2011applied}.

\subsection{Collaborative Outcome}\label{sec:labstudyresults_outcome}
Participants used Blocks to create a total of 48 structures, divided into an equal number of tables, robots, and open-ended structures across all sessions. For the thematic tasks of building tables and robots, participants built a wide range of structures with various styles, color selections, and sizes. They even emulated existing robots such as WALL-E and a Boston Dynamics robot. For the open-ended task, participants selected a diverse set of goals and created structures such as a castle, a sailboat, and a dragon house. Sample creations are shown in Figure~\ref{fig:teaser}.

\begin{figure*}
\centering
\begin{minipage}{.48\textwidth}
 \centering
 \includegraphics[width=\textwidth]{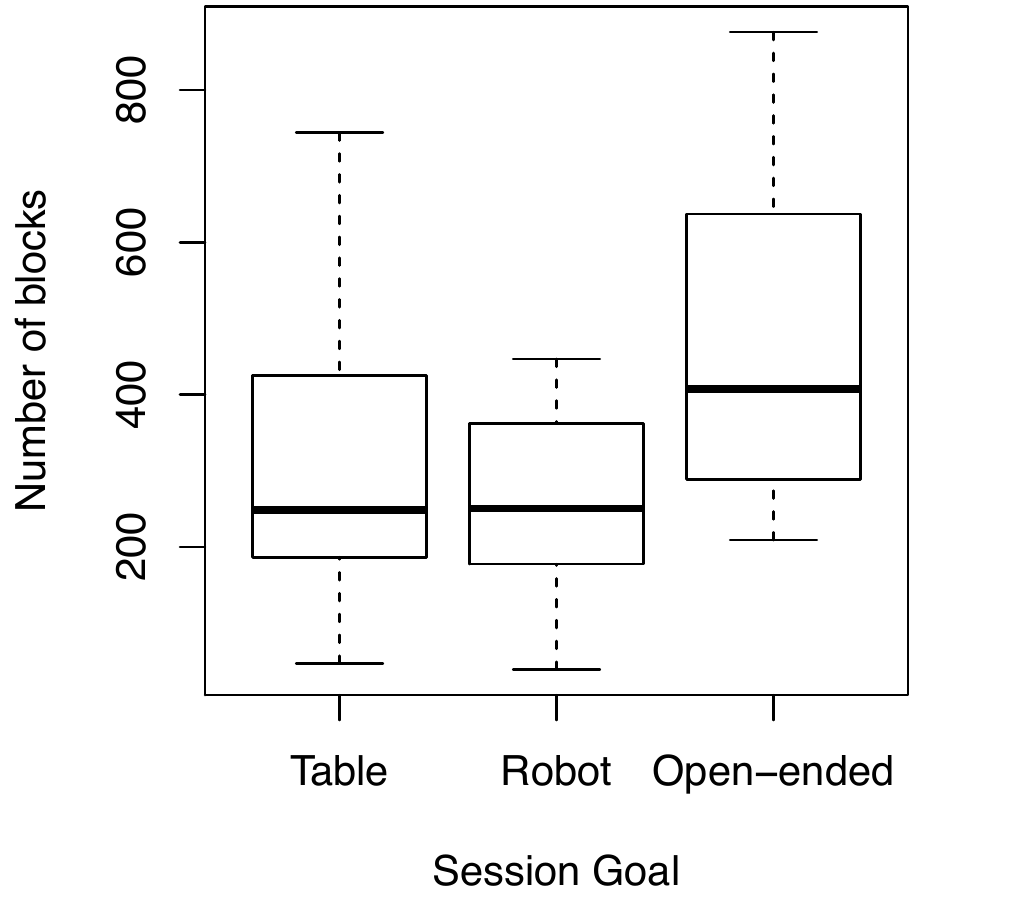}
 \vspace{-1pc}
 \captionof{figure}{Significant main effect of \textit{Session Goal} (tables, robots, open-ended structures) on the number of blocks added was observed.}
 \label{fig:Goal_plot}
\end{minipage}
\hspace{0.3cm}
\begin{minipage}{.48\textwidth}
 \centering
 \includegraphics[width=\textwidth]{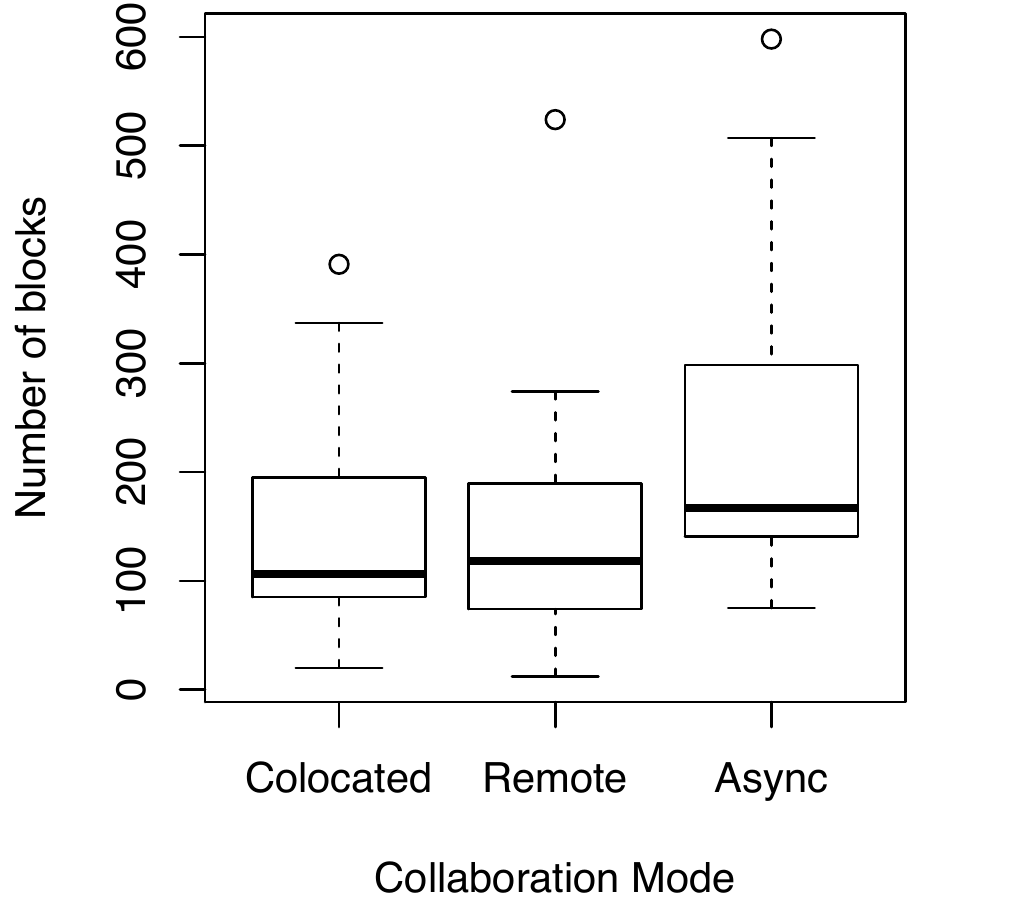}
 \vspace{-1pc}
 \captionof{figure}{Significant main effect of \textit{Collaboration Mode} (colocated-sync, remote-sync, async) on the number of blocks added was observed.}
 \label{fig:NumBlocks_plot}
\end{minipage}
\end{figure*}

We measured the number of blocks added as an indication of the complexity or scale of the structures, similar to the word count in collaborative writing \cite{mabrito2006study, Olson:2017}. We found a significant main effect of \textit{Session Goal} (tables, robots, open-ended structures) on the number of blocks added, \textit{F}(2, 20) = 6.85, \textit{p} = 0.005 (Figure~\ref{fig:Goal_plot}). Additionally, post-hoc analyses revealed that participant pairs added an average of 465 blocks in the open-ended sessions (\textit{SD} = 223), significantly more than in the robots sessions (\textit{M} = 260, \textit{SD} = 129, \textit{t}(17.6) = 2.76, \textit{p} = 0.039) and slightly more than in the tables sessions (\textit{M} = 306, \textit{SD} = 188, \textit{t}(21.4) = 1.89, \textit{p} = 0.144). As mentioned by P14, {\em ``This (open-ended) session was significantly more fun, ... we went from making something very executional like a table, and with totally independent extra elements, to talking out an idea, then building on each other's ideas to come to a creative place neither of us would have gone on our own.''} Participants' self-reported creativity increased from an average of 4.21 (\textit{SD} = 1.10) to 5.46 (\textit{SD} = 0.93) after the study (\textit{t}(44.8) = 4.24, \textit{p} < 0.001).

\subsection{Collaborative Process}\label{sec:labstudyresults_process}
During a typical session, participants first planned their structures, then collaboratively built them, and finally left notes to future creators who might build on top of their creation.

\subsubsection{Planning}\label{sec:labstudyresults_process_planning}
From open-ended responses and study videos, we observed that participants planned their structures verbally, demonstrating with body gestures, and drawing sketches on the whiteboard. They discussed strategies around the styles, shapes, colors, names, number of blocks for each part of their structure, and how they were going to collaborate. We also observed that participants spent less time planning in the async condition compared to colocated-sync and remote-sync conditions. 

\subsubsection{Collaborative Participation}\label{sec:labstudyresults_process_participation}
We found a significant main effect of \textit{Collaboration Mode} (colocated-sync, remote-sync, async) on the number of blocks added, \textit{F}(2, 20) = 5.35, \textit{p} = 0.014 (Figure~\ref{fig:NumBlocks_plot}). Although the differences were not statistically significant at the individual level, post-hoc analyses revealed that participant pairs in the async condition added more blocks (\textit{M} = 454, \textit{SD} = 206) than in the colocated-sync (\textit{M} = 292, \textit{SD} = 167, \textit{t}(21.1) = 2.12, \textit{p} = 0.139) and remote-sync conditions (\textit{M} = 286, \textit{SD} = 192, \textit{t}(21.9) = 2.06, \textit{p} = 0.139).

\begin{figure*}
\centering
\begin{minipage}{.48\textwidth}
 \centering
\vspace{-0.9pc}
 \includegraphics[width=\textwidth]{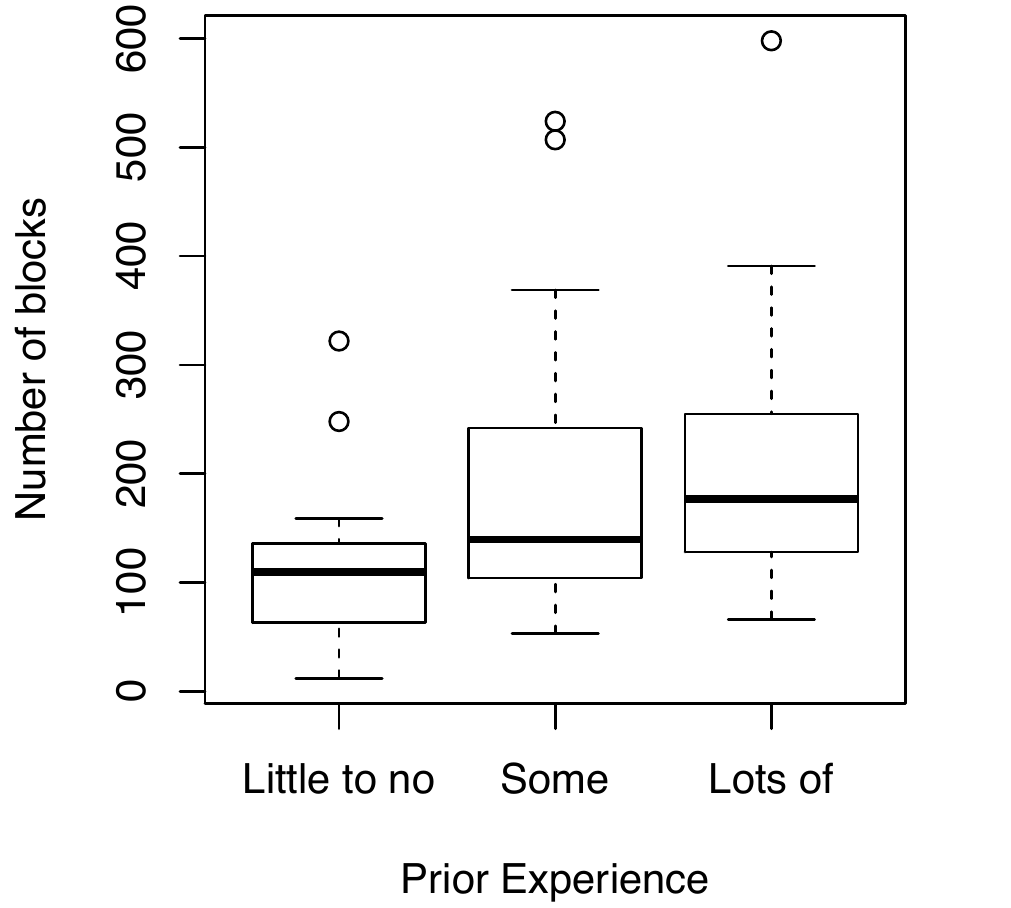}
 \vspace{-1pc}
 \captionof{figure}{Significant main effect of \textit{Prior Experience in AR/VR} (lots of experience, some experience, little to no experience) on the number of blocks added was observed.}
 \label{fig:Experience_plot}
\end{minipage}
\hspace{0.3cm}
\begin{minipage}{.48\textwidth}
 \centering
 \includegraphics[width=\textwidth]{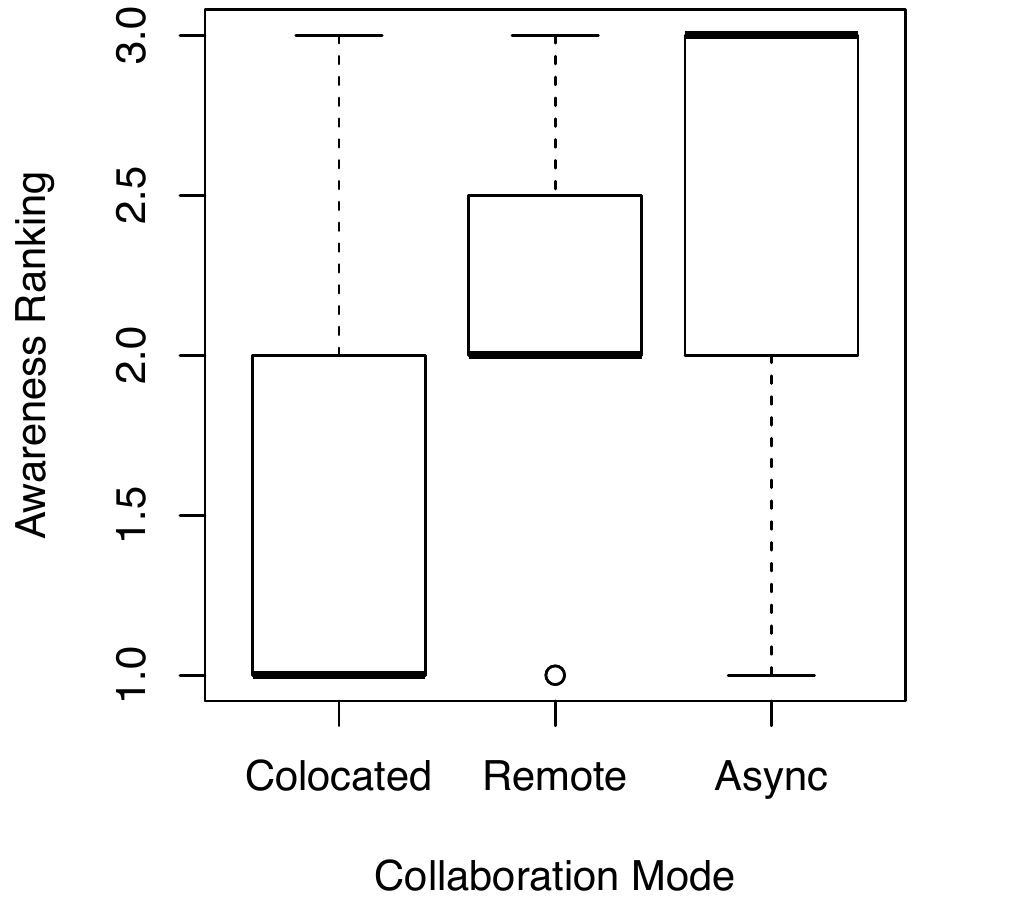}
 \vspace{-1pc}
 \captionof{figure}{Significant main effect of \textit{Collaboration Mode} (colocated-sync, remote-sync, async) on the ranking about the ease to be aware of their partners' actions was observed (1 being the best rank).}
 \label{fig:Awareness_plot}
\end{minipage}
\vspace{-1pc}
\end{figure*}

Although the differences were not statistically significant, we observed that participants deleted each others' work less in the async condition (\textit{M} = 0.44\%) than in the colocated-sync (\textit{M} = 2.83\%) and remote-sync (\textit{M} = 2.82\%) conditions (\textit{t}(11.4) = 1.69, \textit{p} = 0.118). This might be related to participants being concerned about deleting their partner's work without being able to communicate with them.

As shown in Figure~\ref{fig:Experience_plot}, we also found a main significant effect of self-reported \textit{Prior Experience in AR/VR} on the number of blocks added (\textit{F}(2, 67) = 4.93, \textit{p} = 0.010), i.e., ``experts'' (\textit{M} = 207, \textit{SD} = 128) added more blocks than participants with ``some'' experience (\textit{M} = 190, \textit{SD} = 123) and with ``little to no experience'' (\textit{M} = 111, \textit{SD} = 74).

\subsubsection{Participation Balance}\label{sec:labstudyresults_process_balance}
We measured how balanced participants' contributions were (measured by the number of blocks added) because this can reveal skewed contributions, leadership, and other roles that can affect end-user experiences and credit distribution \cite{crowdresearch}. Inspired by the study of participation in collaborative writing \cite{Olson:2017}, we calculated a measure of {\em participation balance} as one minus the variance of the individual contribution percentages. For example, if the two participants contributed equally, the {\em participation balance} would be 1.0. Similarly, if only one person contributed, the balance would be 0.5. In our lab study, we observed that the participation balance was high (\textit{M} = 0.960, \textit{SD} = 0.057) and consistent across colocated-sync (\textit{M} = 0.962, \textit{SD} = 0.081), remote-sync (\textit{M} = 0.958, \textit{SD} = 0.029), and async (\textit{M} = 0.961, \textit{SD} = 0.055) conditions. Furthermore, in survey responses, participants also expressed feeling that both their partner and themselves contributed substantially to the outcome (\textit{M} = 6.03, \textit{SD} = 1.10), which is consistent with the log data collected during these sessions.

\subsubsection{Communication and Coordination}\label{sec:labstudyresults_process_communication}
Across all conditions, participants strongly agreed that it was easy to work with each other (\textit{M} = 5.96, \textit{SD} = 1.00). We also asked participants to compare how easy it was to be aware of their partners' actions across the three conditions and we found a significant main effect of the three conditions on the rankings, \textit{$\chi^2$}(2) = 11.58, \textit{p} = 0.003 (Figure~\ref{fig:Awareness_plot}). Post hoc analyses revealed that participants ranked the colocated-sync condition significantly higher than the remote-sync (\textit{Z} = -2.44, \textit{p} = 0.030) and async conditions (\textit{Z} = -2.64, \textit{p} = 0.026). In open-ended responses, participants commented that it was hard to know what their partner was doing in the remote-sync condition without seeing each other (P4, P21), and it was even harder to build together in the async condition without verbal communication (P1, P15). This indicates that colocated-synchronous collaboration enabled better communication and coordination.

We also found a significant main effect of \textit{Collaboration Mode} on the number of colors participants used, \textit{F}(2, 20) = 5.88, \textit{p} = 0.010 (Figure~\ref{fig:Colors_plot}). Post-hoc analyses revealed that participants used significantly more colors in the async condition (\textit{M} = 11.8, \textit{SD} = 6.62) than in the colocated-sync condition (\textit{M} = 5.75, \textit{SD} = 3.22, \textit{t}(15.9) = 2.82, \textit{p} = 0.037), although the differences between the colocated-sync and remote-sync conditions (\textit{M} = 7.25, \textit{SD} = 7.21) were not statistically significant (\textit{t}(21.8) = 1.59, \textit{p} = 0.251). This difference of color usage might be due to the limited communication between the participants in the async condition. For example, P6 left a note for future creators to use the color picker to choose the same color, hinting to the fact that limited communication might result in inconsistent colors from multiple collaborators.

\subsubsection{Physical Movement}\label{sec:labstudyresults_process_movement}
People build structures with Blocks in a way similar to how one would play with LEGO bricks in the physical world. For example, we observed participants moving around, kneeling down, lying on the ground, and standing on the chairs, in order to get closer to the parts they were building. In open-ended responses, P16 said: {\em ``if you build very tall, you may need to use a chair to keep building on top!''} Interestingly, we observed in the study videos that participants walked less in the colocated-sync condition than in the other two conditions. This might be because colocated participants were worried about getting into their partner's way as they are focusing on the creation. As noted by P10, {\em ``I was worried that I might be in the way.''}

\subsubsection{Notes for Future Creators}\label{sec:labstudyresults_notes}
Since one of the features of Blocks is that structures stay persistent for others to see and change, we wanted to understand participants' creative visions and their perceptions of ownership. To do so, we asked participants to leave a note behind for future creators. We organized the notes in three non-mutually exclusive categories: (i) \emph{instructions} on how to make changes, (ii) \emph{advice} on creation techniques, and (iii) \emph{ownership} signals by the original creator. For notes that left instructions for the next creator, 26\% asked to add components to their creation, 30\% asked to refine what they had created, and 16\% encouraged the next person to be creative, i.e., do whatever they want. For notes that provided advice, 21\% gave guidance on using the app itself, and 4\% on how to interact with their partner more effectively. Lastly, for notes related to ownership, 13\% of the notes asked the next person to preserve what they had built, and 17\% articulated their vision for the creation. The results show some participants wanted others to modify their work and others did not, hinting at the need for future permission control mechanisms.

\begin{figure*}
\centering
\begin{minipage}{.38\textwidth}
 \centering
 \includegraphics[width=\textwidth]{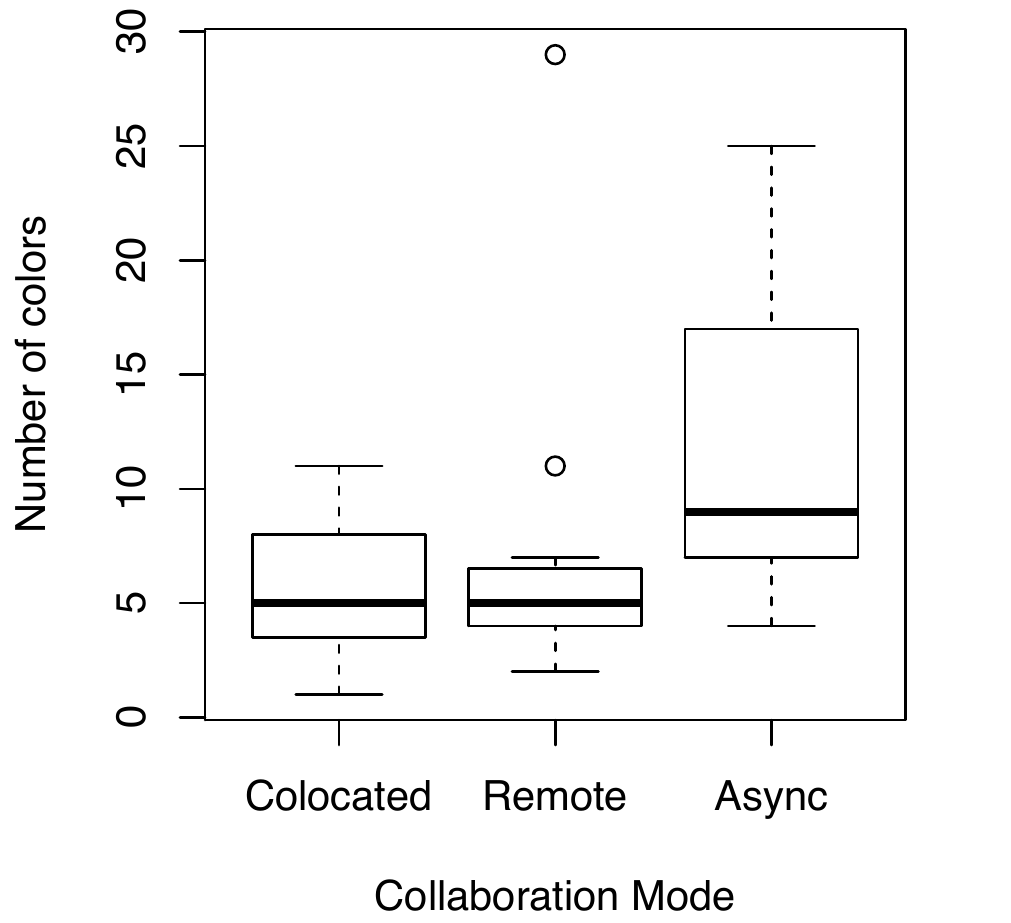}
 \vspace{-1pc}
 \captionof{figure}{Significant main effect of \textit{Collaboration Mode} (colocated-sync, remote-sync, async) on the number of colors used was observed.}
 \label{fig:Colors_plot}
\end{minipage}
\hspace{0.3cm}
\begin{minipage}{.58\textwidth}
 \centering
 \vspace{-.6pc}
 \includegraphics[width=\textwidth]{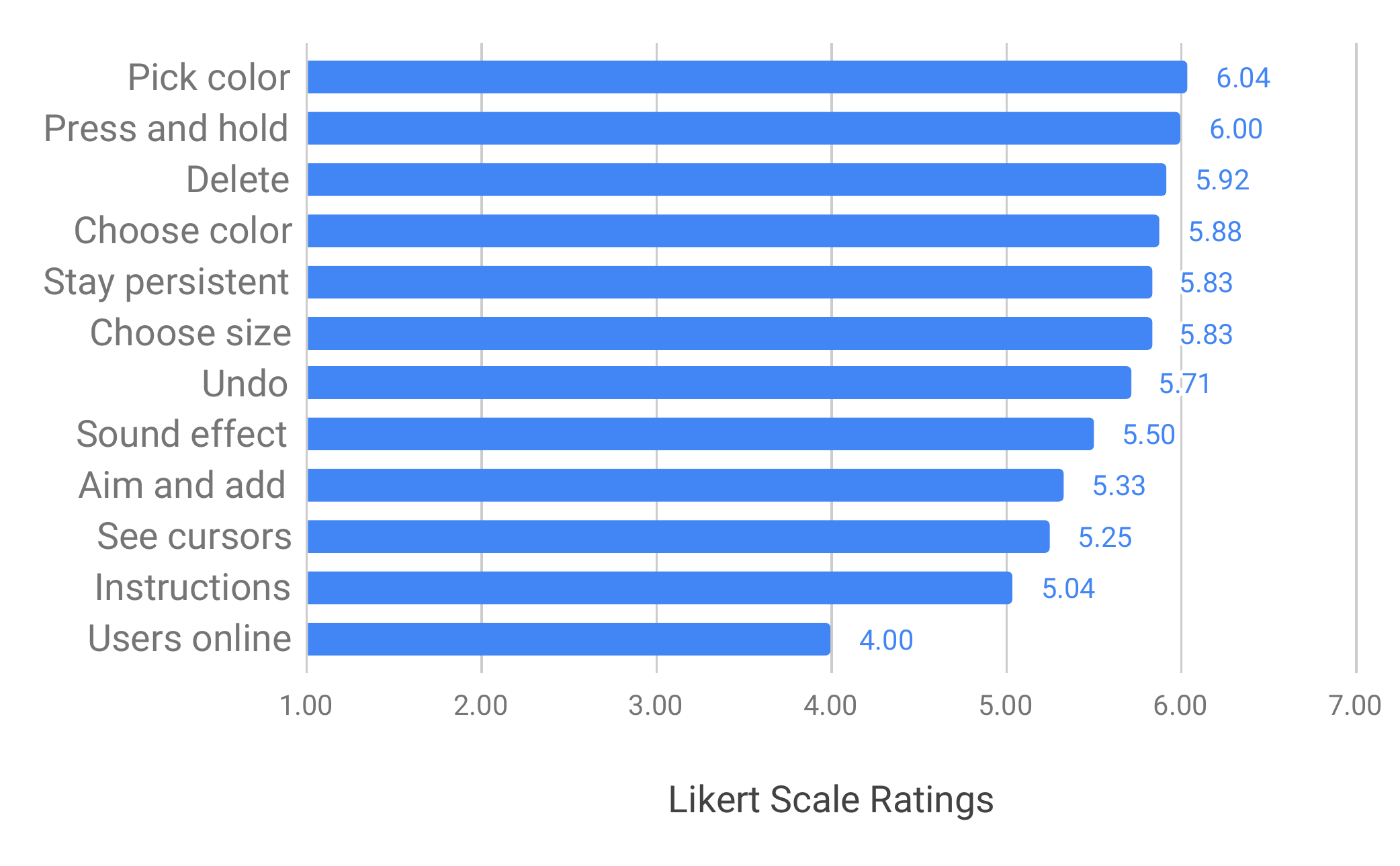}
 \vspace{-1.7pc}
 \captionof{figure}{Participants' Likert scale ratings on the usefulness of the various features of Blocks (7 being extremely useful).}
 \label{fig:FeatureRatings}
\end{minipage}
\end{figure*}

\subsection{User Experience and Feedback}\label{sec:labstudyresults_ux}

\subsubsection{Usefulness of Features}\label{sec:labstudyresults_ux_features}
We asked participants to rate the usefulness of the core features of Blocks. We summarize the results in Figure~\ref{fig:FeatureRatings}. Participants extremely agreed that the ability to pick the color of an existing block was useful (\textit{Md} = 7, \textit{M} = 6.04, \textit{SD} = 1.52). Participants strongly agreed (\textit{Md} = 6) on the usefulness of the following features including the ability to aim and tap to add a block (\textit{M} = 5.33, \textit{SD} = 1.55), press and hold to keep adding blocks (\textit{M} = 6.00, \textit{SD} = 1.25), see others cursors in real time (\textit{M} = 5.25, \textit{SD} = 1.48), choose colors (\textit{M} = 5.88, \textit{SD} = 1.23) and sizes (\textit{M} = 5.83, \textit{SD} = 1.43), undo their own blocks (\textit{M} = 5.71, \textit{SD} = 1.46), delete any blocks (\textit{M} = 5.92, \textit{SD} = 1.28), as well as the sound effects for adding and deleting blocks (\textit{M} = 5.50, \textit{SD} = 1.53), and that the blocks added will stay persistent for others to see and edit (\textit{M} = 5.83, \textit{SD} = 1.37). These results suggest that Blocks' creation tools and interactions were effective.

On the other hand, participants only somewhat agreed that the instructional hints were useful (\textit{Md} = 5, \textit{M} = 5.04, \textit{SD} = 1.55). This could be because the single-screen onboarding overlay might not have provided enough scaffolding for the individual features. Participants were neutral (\textit{Md} = 4) on the usefulness of the counter of users online (\textit{M} = 4.00, \textit{SD} = 1.87). We suspect this is because there were only two participants using Blocks during the lab study, but might be more useful with more concurrent users.

\subsubsection{Happiness, Creativity, and Closeness}\label{sec:labstudyresults_ux_happy}
Participants reported feeling happier, more creative, and closer with their partners after using Blocks. 

Participants' self-reported happiness showed a significant increase from an average of 5.17 (\textit{SD} = 0.82) to 5.79 (\textit{SD} = 1.02) after the study (\textit{t}(43.9) = 2.34, \textit{p} = 0.012). For instance, P9 mentioned that \emph{``after a long day at work it's fun to play a game that is engaging and work with other people to complete challenges.''}
Furthermore, P13 mentioned experiencing different kinds of happiness and energy when collaborating synchronously vs. asynchronously: 
\quotepar{``I think it's a different kind of happiness. The first (sync) -- the moment you're working together and building on each others' it is just fun and interesting. And when working on something and then switch (async), that fun part is on the result part. When you see the outcome, and wow!''}{(P13)}

Participants' self-reported creativity also increased from an average of 4.21 (\textit{SD} = 1.10) to 5.46 (\textit{SD} = 0.93) after the study (\textit{t}(44.8) = 4.24, \textit{p} < 0.001). For example, P14 mentioned how their {\em ``creativity and fun were heightened by working with (their) partner.''} Interestingly, P13 felt that her creativity decreased: {\em ``my partner is way better than me in creation and building.''}

Participants' self-reported closeness with their study partners also increased from an average IOS score of 3.04 (\textit{SD} = 1.55) to 4.5 (\textit{SD} = 1.14) after the study (\textit{t}(42.3) = 3.82, \textit{p} < 0.001).

\quotepar{``I feel more than we even audibly communicated, we got a sense of each others creative rhythms and we made something together that was more fun and interesting and creative than had we worked independently. I feel I've gained something by working with her and that brings a closeness. I don't know any more about my partner than when we started but I feel more connected and comfortable with her.''}{(P14)}

\begin{figure*}
\centering
\begin{minipage}{.48\textwidth}
 \centering
 \includegraphics[width=\textwidth]{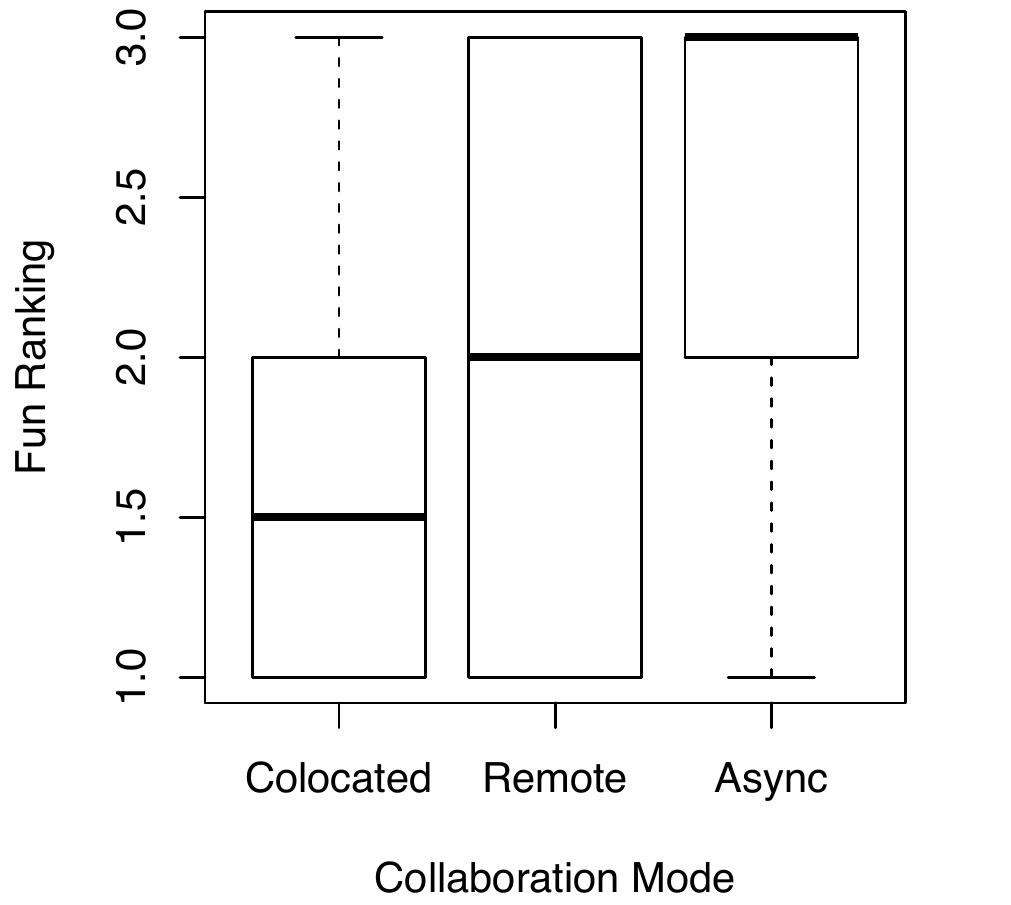}
 \vspace{-1pc}
 \captionof{figure}{Significant main effect of \textit{Collaboration Mode} (colocated-sync, remote-sync, async) on the ranking about having fun was observed (1 being the best rank).}
 \label{fig:Fun_plot}
\end{minipage}
\hspace{0.3cm}
\begin{minipage}{.48\textwidth}
 \centering
 \includegraphics[width=\textwidth]{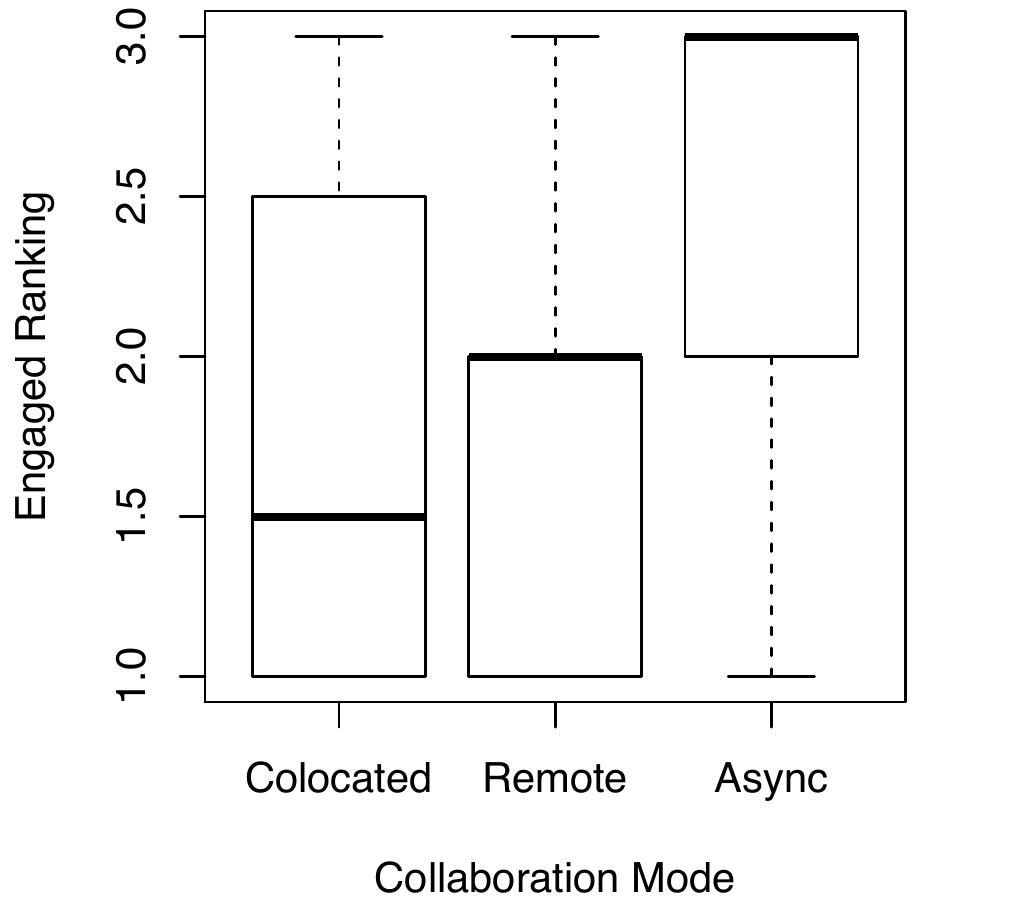}
 \vspace{-1pc}
 \captionof{figure}{Significant main effect of \textit{Collaboration Mode} (colocated-sync, remote-sync, async) on the ranking about being engaged was observed (1 being the best rank).}
 \label{fig:Engaged_plot}
\end{minipage}
\vspace{-0.6pc}
\end{figure*}

\subsubsection{Fun, Engagement, and Satisfaction}\label{sec:labstudyresults_ux_fun}
Participants strongly agreed that they had fun (\textit{M} = 6.33, \textit{SD} = 1.14) and feeling engaged (\textit{M} = 6.36, \textit{SD} = 1.00) across all three conditions. Participants ranked the conditions in the following order: colocated-sync, remote-sync, and async for both fun (\textit{$\chi^2$}(2) = 6.08, \textit{p} = 0.048) and engagement \textit{$\chi^2$}(2) = 9.00, \textit{p} = 0.011), as shown in Figure~\ref{fig:Fun_plot} and \ref{fig:Engaged_plot}.

\quotepar{``It was fun even before we began building to just discuss what we wanted to make, then watch how it evolved as we began creating. I really enjoyed watching it come together and each of us adding small elements because we both didn't care if we strayed from the original idea of how the robot should appear.''}{(P10)}

\quotepar{``It's actually deceptively simple and the concept is so obvious as it's super familiar to real life playing with blocks/LEGO. I still have never seen this execution before and it's really fun and remarkable.''}{(P14)}

In survey responses, participants strongly agreed feeling satisfied with their creation (\textit{M} = 6.10, \textit{SD} = 1.08), and that they were able to accomplish their goal (\textit{M} = 6.32, \textit{SD} = 0.96).
\quotepar{``This was a fun session! I loved collaborating with my partner and having the final product be something we built together..., the ultimate payoff of building what we built was awesome!''}{(P16)}

Participants' feedback indeed skewed positive and were supported by open-ended comments. We believe the feedback was genuine because participants also wrote critical feedback such as technical difficulties (e.g., the app crashing and network connectivity issues).

\quotepar{``Really fun and engaging experience. I think this has a lot of potential value in the AR realm, and the interactivity of the tasks made it much more fun than working alone. There were a few bugs, and as the designs grew, it got slower and slower, but overall the experience was great.''}{(P9)}

Furthermore, participants often continued working on their creations for longer than we asked them to (15 minutes), spending an average of 16.8 minutes per session (\textit{SD} = 5.88, 17.0 minutes for colocated-sync, 14.5 minutes for remote-sync, and 18.8 minutes for async). In one colocated-sync session, for example, participants spent more than 30 minutes crafting a 3D structure until they were satisfied with it, showing that the experience was highly engaging. The time spent in the async condition was marginally longer than that in remote-sync condition (\textit{t}(15.9) = 2.50, \textit{p} = 0.071). 

\subsection{Summary of Lab Study Results}\label{sec:labstudyresults_summary}
Our lab study aimed to evaluate the effectiveness of Blocks in supporting end-user creation, and how people's location affects their creation and collaboration. We learned that participants felt engaged, more creative, and closer to their partners while using Blocks. They created a wide variety of structures, from tables to castles, from robots to sailboats (see Figure~\ref{fig:teaser}). In terms of the quality of the experience, participants reported enjoying synchronous colocated collaboration the most. Overall, we found it promising that many participants in the lab study willingly spent more time working on their creations than required.

However, due to the nature of a lab study, it is unclear how participants would engage with Blocks in a naturalistic setting. Furthermore, while we could study the effect of \textit{people}'s location in a lab set up in short sessions, we still lacked the understanding of how \textit{AR structures}' location affects creation and collaboration (Table~\ref{tab:DesignSpace}). We could best study these in a setting that reflects people's regular routine over longer time spans. Therefore, we deployed Blocks in the wild.

\section{Field Study}\label{sec:fieldstudy}
The goal of our field study was two-fold: (i) to evaluate how end users engage with Blocks in a naturalistic setting, and (ii) to investigate how creation and collaboration differ when the AR structures are location-dependent vs. independent, as indicated in Table~\ref{tab:DesignSpace}.

\subsection{Participants and Method}
\begin{figure}[t!]
 \centering
 \includegraphics[width=0.8\linewidth]{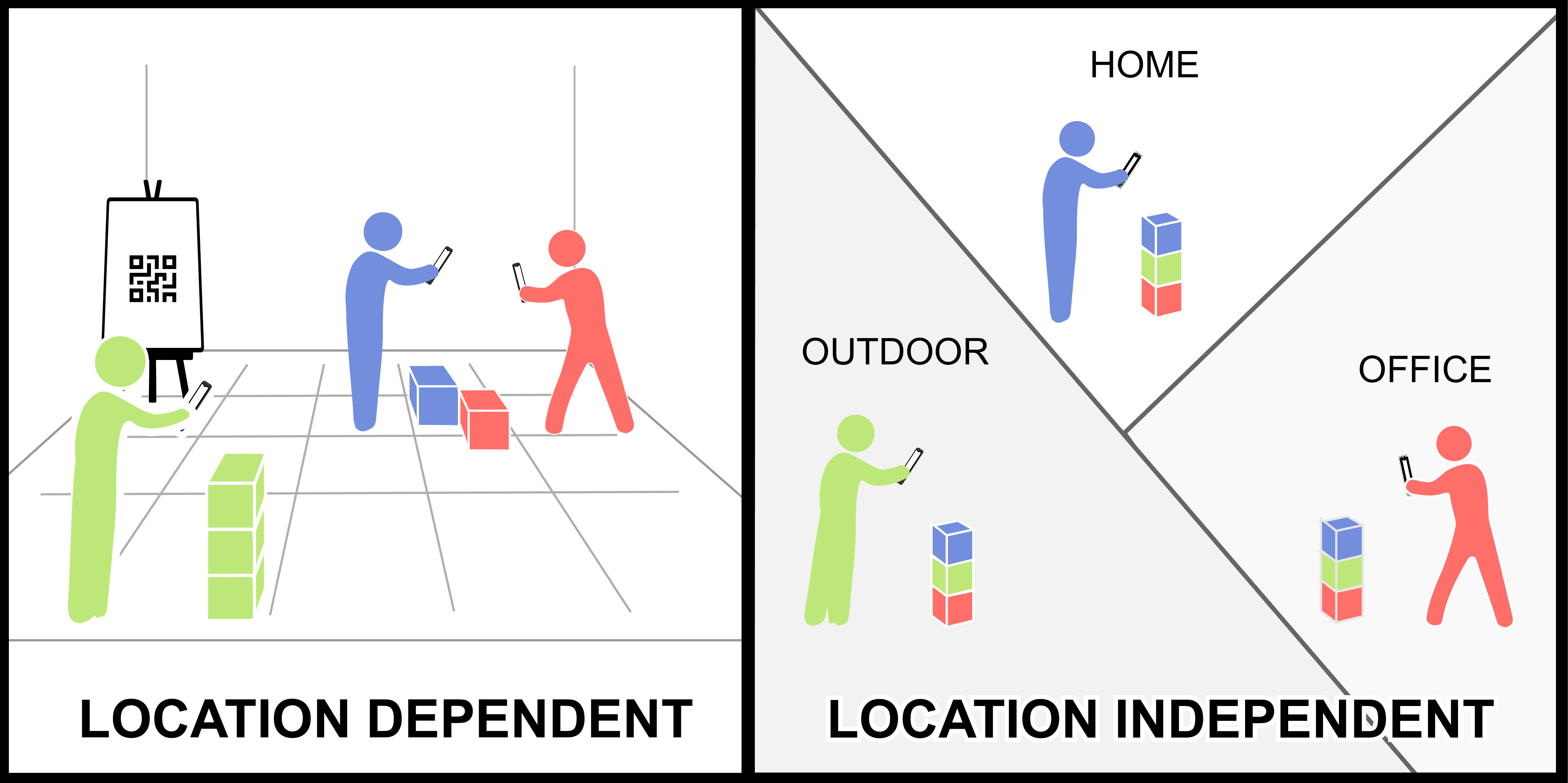}
 \caption{Experimental setup for the field study. In the \textit{location-dependent} scenario (1), the AR structures could only be accessed at the poster's location. In the \textit{location-independent} scenario (2), the same AR structures could be accessed anywhere at any time.}
\label{fig:FieldSetup}
\end{figure}

We recruited 68 participants throughout a three-day field study. We recruited these participants by setting up posters at two visible locations: the lobby of an office building, and a cafeteria's outdoor seating area. We conducted the studies at the lobby on Days 1 and 2 and at the cafeteria on Day 3. On Day 1, we deployed the \textit{location-dependent} scenario, where participants had to return to the poster's location to access the AR structures (Figure~\ref{fig:FieldSetup}.1). On Day 2, we deployed the \textit{location-independent} scenario was deployed where participants could access the AR structures from anywhere at any time, i.e., users would see the structures overlaid on their physical environment (Figure~\ref{fig:FieldSetup}.2). On Day 3, we deployed the location-dependent scenario in the morning and the location-independent scenario in the afternoon.

\subsection{Procedure and Setup}
When participants walked up to a poster, they first scanned a code to install the Blocks iOS app. Participants first tried out Blocks in their personal sandbox (``MyWorld''), then entered the shared AR world (``OurWorld'') to view what others have created and to create anything they wanted others to see. In the location-dependent scenario, Blocks used the poster itself as the marker for maintaining location-dependence. Before the participants left, they noted down their email address and in-app identifier in a Google Form, which we used to solicit post-study survey responses. We also stored activity logs on the server for further analysis.

\section{Field Study Results}\label{sec:fieldstudyresults}
Across three days, a total of 68 participants used Blocks during 222 sessions, and added a total of 6,970 blocks in both MyWorld and OurWorld. 

\begin{table*}
\caption{Descriptive statistics from our field study, including general results, creation, collaboration, and post-study survey subjective ratings in both location-dependent and independent scenarios.}
\label{tab:DeploymentStats}
\vspace{-.4pc}
 \renewcommand{\arraystretch}{1.1}
\centering
\footnotesize
\begin{tabular}{l|l|l}
 & \textbf{Location-Dependent} & \textbf{Location-Independent} \\
 \specialrule{.1em}{.05em}{.05em} 
\textbf{General Statistics} & & \\
Users & 32 & 36 \\
User sessions & 91 & 131 \\
Users with \textgreater{}1 session & 22 & 28 \\
Sessions with blocks added & 65 & 85 \\
Avg blocks per session & 58 & 35 \\
Avg time per session (min) & 4.2 & 5.8 \\ \hline
\textbf{OurWorld} & & \\
Blocks added & 1,722 & 1,936 \\
Users & 25 & 26 \\
Blocks deleted & 96 & 438 \\
Deletion by others & 14 & 93 \\ \hline
\textbf{Collaboration} & & \\
Synchronous moments & 15 & 16 \\
Sync: blocks added & 490 & 238 \\
Asynchronous sessions & 76 & 115 \\
Async: blocks added & 1,232 & 1,698 \\ \hline
\textbf{Ratings (1-7): Mean (SD)} & & \\
``I like that AR structures \textit{persist} only at that location vs. anywhere'' & 6.31 (1.20) & 6.24 (1.15) \\
``I like that I could \textit{add} only at that location vs. anywhere'' & 5.31 (2.12) & 6.47 (0.87) \\
``I like that I could \textit{collaborate} only at that location vs. anywhere'' & 5.75 (1.57) & 6.29 (1.05) \\
``I like that I could \textit{view} only at that location vs. anywhere'' & 5.81 (1.68) & 6.24 (1.09) \\
``Blocks is fun to use'' & 6.18 (0.87) & 5.50 (1.45) \\
``Blocks is engaging'' & 6.64 (0.50) & 5.83 (1.34)
\end{tabular}
\vspace{-.6pc}
\end{table*}

\subsection{Collaborative Outcome and Process}\label{sec:fieldstudy_outcomeprocess}
Participants created many styles of trees, towers, windows, tables, stairs, hearts, and word art. Participants also mentioned playing with Blocks for various purposes, such as trying to create {\em ``the highest tower with continuous blocks''} (P59) and competing with other players, e.g., P65 wanted to {\em ``build a tower taller than [P64]'s.''} Overall, participants spent an average of 5.1 minutes in each session.\footnote{We defined a ``session'' as opening and closing the Blocks app. For session duration, we only measured the ones that involved adding a block.}

\begin{figure*}[t!]
\centering
\begin{minipage}{.48\textwidth}
 \centering
 \includegraphics[width=\textwidth]{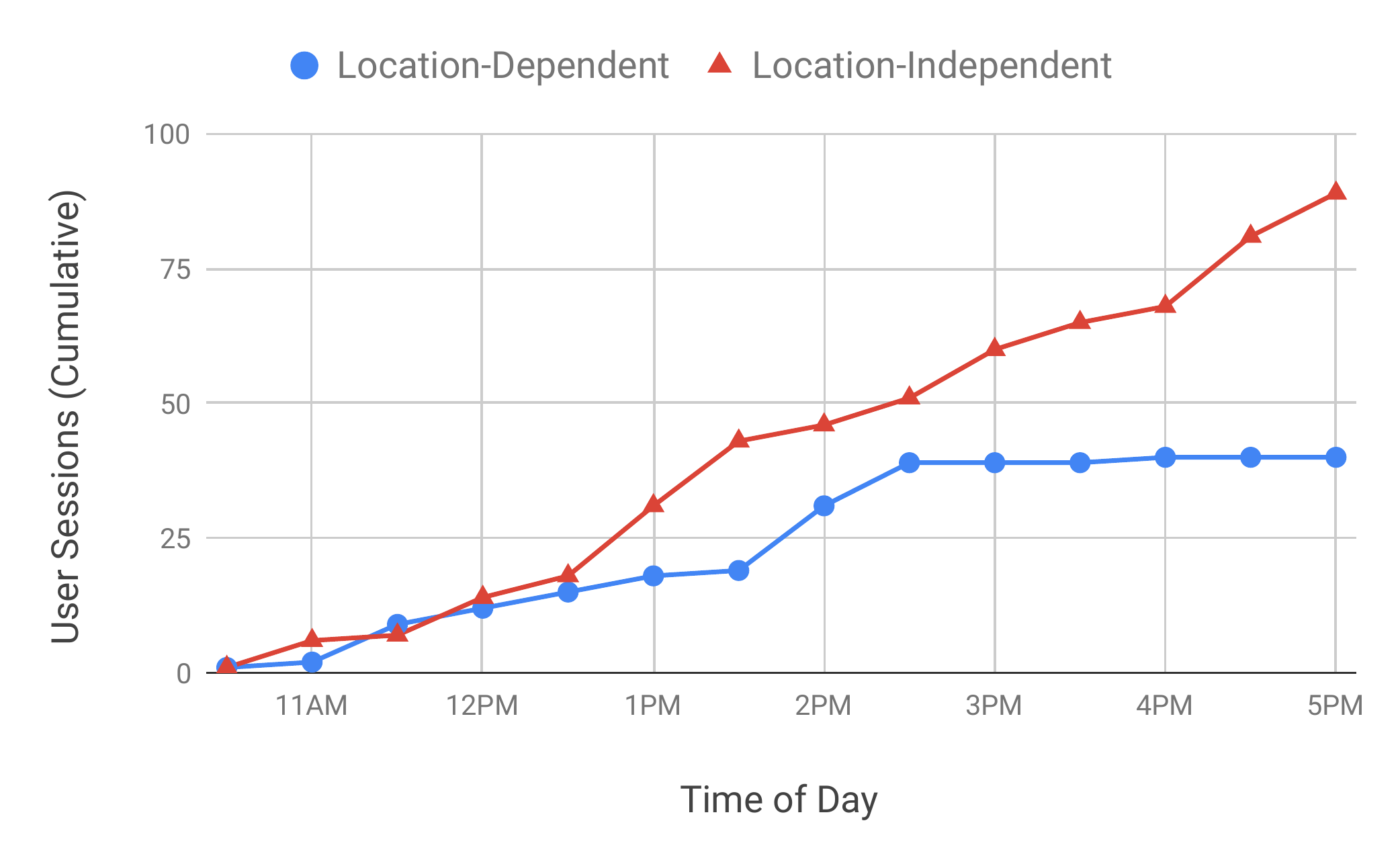}
 \vspace{-1.7pc}
 \captionof{figure}{Cumulative number of user sessions throughout the day for location-dependent and independent scenarios. Note that participants engaged in more sessions during the location-independent scenario.}
 \label{fig:TimelineSession}
\end{minipage}
\hspace{0.3cm}
\begin{minipage}{.48\textwidth}
 \centering
 \includegraphics[width=\textwidth]{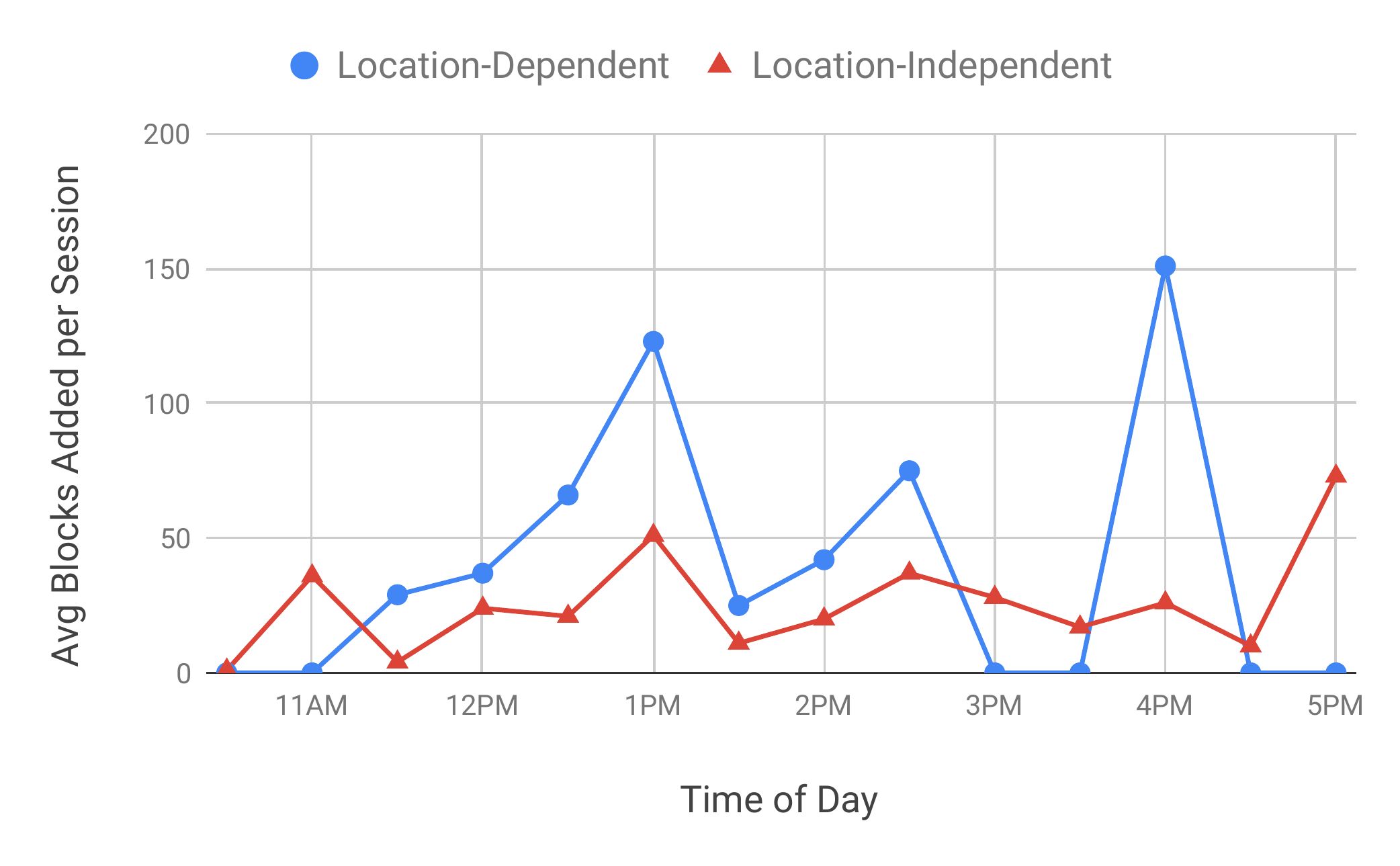}
 \vspace{-1.7pc}
 \captionof{figure}{Average number of blocks added throughout the day for location-dependent and independent scenarios. Note that when the AR structures were location-dependent, participants added more blocks per session on average.}
 \label{fig:TimelineBlocks}
\end{minipage}
\end{figure*}

As shown in Table~\ref{tab:DeploymentStats}, we had 32 participants using Blocks in 91 sessions in the location-dependent scenario. The average session, in this scenario, lasted 4.2 minutes and had 58 blocks added by participants. Furthermore, participants added at least one block in 65 of the 91 sessions. Lastly, we found that 22 participants (69\%) used Blocks more than once during a given day of the study. 

In the location-independent scenario, 36 participants accessed Blocks in 131 user sessions. The average session, in this scenario, lasted 5.8 minutes and had 35 blocks added by participants. Furthermore, participants added at least one block in 85 of the 131 sessions. Lastly, we found that 28 participants (78\%) repeatedly used Blocks.

As we show in Figure~\ref{fig:TimelineSession} and ~\ref{fig:TimelineBlocks}, when the AR structures were location-dependent, i.e., when participants had to return to the previous location to access them again, participants added more blocks but spent less time in each user session. This was possibly due to users' limited time availability when passing by that specific location (e.g., note how there are almost no new sessions after 3pm). Participants also engaged in fewer sessions because of the location constraints. On the other hand, when the AR structures could be accessed anywhere at any time, more users accessed the app in more sessions, indicating high engagement and activity. Overall, 74\% of participants used Blocks repeatedly during the deployment. 

Participants collaborated both synchronously and asynchronously. We identified 31 moments when multiple users were collaborating synchronously\footnote{We defined synchronous collaboration as a moment when two or more users added blocks simultaneously.} with each other, either colocated or remote. The rest of the user sessions were treated as asynchronous collaboration moments in which participants created new structures or added on top of existing ones to build a common ``OurWorld.''

In the location-dependent scenario, we identified 15 synchronous collaboration moments (colocated-sync), resulting in a total of 490 blocks added, compared to 1,232 blocks added through asynchronous collaboration during 76 sessions. On the other hand, for the location-independent scenario, 16 synchronous collaboration moments (remote-sync) were identified, resulting in a total of 238 blocks added, compared to 1,698 blocks added through asynchronous collaboration during 115 sessions. This indicates that participants were more active when they created structures at the time and place of their choosing.

In multiple scenarios, two participants were initially collaborating synchronously at the poster's location. As people passed by, they were curious to see what the two were building, then joined their collaboration sessions. Through these multi-user collaborative experiences, participants received diverse feedback from their peers, and easily engaged in discussions and interactions, which were one of the highlights of the field study.

\subsection{User Preferences}\label{sec:fieldstudy_ux}
A total of 28 participants responded to our post-study survey: 11 respondents participated in only the location-dependent scenario, 12 respondents participated in only the location-independent scenario, and the other 5 participated in both conditions. As in the lab study, participants answered 7-point Likert scale questions, where 1 was extremely negative and 7 was extremely positive. For both location-dependent and independent scenarios, we asked questions focusing on the impact of location on experiences of consumption (they could \textit{view} only at that location vs. anywhere), and co-creation (they could \textit{add/collaborate} only at that location vs. anywhere). In both conditions (Table~\ref{tab:DeploymentStats}), participants' ratings were high, with participants rated the location-independent scenario slightly higher regarding adding, collaborating and viewing. For participants who participated in both conditions, we asked them to rank the two. Their preferences for the conditions were balanced, and a case could be made for the benefits of each condition.
\quotepar{``I think the location specific idea of these blocks adds a magical element of there being a persistent world taking place in parallel to what you see all day.'' }{(P27)}
\quotepar{``Creativity happened to strike while I was at my desk and that was the moment I wanted to create.''}{(P30)}

Participants strongly agreed that Blocks was fun to use (\textit{M} = 5.93, \textit{SD} = 1.18), and they felt strongly engaged (\textit{M} = 6.25, \textit{SD} = 1.00). Furthermore, 74\% of the users accessed Blocks more than once during the study.

\subsection{Effect of Physical Environments}\label{sec:fieldstudy_physicalenv}
Physical environments played an important role in the participants' creation. Some participants made creations that interacted with the real world. For example, P46 built a little house next to their desk and P62 built on top of a fire pit to add a virtual fire.

\quotepar{``I used the walkway between desks as the boundary of where I was building. I did that subconsciously, though. It just seemed like a good place to build!''}{(P30)}
\quotepar{``At my desk, I added a `bridge' that people were unknowingly walking through. Pretty cool.''}{(P30)}

In particular, for the location-independent scenario where participants could use Blocks anywhere, some were limited by their in-the-moment environments.
\quotepar{``When I started making art, I was in a big room, and it was a little more difficult to use when I got back to my desk and had more objects around.''}{(P42)}

Similarly, for the location-dependent scenario, the choice of location became a limiting factor for some. 
\quotepar{``I do think that we were in a small corner too so it would be interesting to do it in a bigger place with more objects so I can build things around them. For example, if there was a cookie jar, I could build a jail cell around it so it can look like the cookies are on lock down.''}{(P55)}

\subsection{Effect of Prior Creations in the Shared AR World}\label{sec:fieldstudy_prior}
Others' creations influenced the users in various ways. On the positive side, participants found that it was fun and inspiring to build on what others have created, and they liked to discover that others had been using Blocks as well. For example, P59 noted {\em ``It was fun to add to what other people had created''}, and P27 mentioned {\em ``For me I got ideas based on what other people had made.''}

On the other hand, participants mentioned being negatively affected by others' creations, especially as it cluttered space over time.
\quotepar{``I feel like because there were so many blocks, it would get a little glitchy when I was up close, but standing farther from the structure worked out well.''}{(P39)}
\quotepar{``I tried to work outside the bounds of what had already been created.''}{(P71)}

Interestingly, we found that courtesy and etiquette among users formed immediately in the shared AR world. For example, several participants mentioned removing random blocks in the scene, but no one removed anything that they perceived as a structure. This is consistent with responses from the lab studies in which participants described not wanting to get in the way when their partners were trying to build. In their note to future creators, many participants from the lab study asked that others not remove or substantially alter their creations.
\quotepar{``Other structures influenced my work, for sure. I didn't want to build in a way that obscured the previous vision. I tried to make it complementary or separate.''}{(P30)}

\subsection{Summary of Field Study Results}\label{sec:fieldstudyresults_summary}
Our field study aimed to evaluate how end users engage with Blocks in a naturalistic setting, and how AR structures' location affects end-user participation. We observed that when participants had to come to a particular location to create structures (location-dependent), they added more blocks per session. On the other hand, when they could create structures anywhere at anytime (location-independent), they participated in more sessions throughout the day. This indicates that location dependence plays an important role in terms of engagement and activity. Overall, many participants in the field study repeatedly engaged with Blocks in a naturalistic setting. Our findings can help AR designers make informed decisions while creating experiences to encourage high engagement and activity.

\section{Extended Deployment}\label{sec:deployment}
After conducting the lab and field studies, we integrated popular user feedback around onboarding, cursor style and scattered blocks, and enabled users to experience Blocks across all the design dimensions for an extended period of time.

We deployed Blocks through a five-day interactive installation set up concurrently at two office building lobbies during the 2018 winter holiday season. This installation consisted of a poster and two large monitors, one that connected with the other building through a continuous video and audio conferencing call, and another that showed live creations. This setup enabled colocated and remote collaboration across the buildings. Also, it allowed people to communicate both visually and verbally. Furthermore, people could co-create structures at the location of the installation or experience them anywhere at any time.

We featured a different holiday theme each day (e.g., ``snowmen'' and ``Christmas trees'') to scaffold people's creativity. Participants could follow the theme of the day or create anything they wanted. The rest of the study procedure was the same as the field study.

Across the five days, a total of 70 participants interacted with Blocks during 447 sessions. Collectively, participants spent 15.1 hours in the app, and added a total of 8,885 blocks. While most participants were inspired by the holiday themes and created many styles of snowmen, Christmas trees and Santa Clauses, others happily steered off track with extensive sculptures of race cars, the Hollywood Sign, a menorah, velociraptors, and DNA molecules. We observed a greater variety of creations than in the field study. In 199 of the 477 sessions (45\%), participants added at least one block, while the rest 55\% were consumption-only sessions. The longest session lasted about 24 minutes, with 303 blocks added to build a race car track. As we had previously mentioned, we found that participants voluntarily spent more time using Blocks than it was needed for the study.

Furthermore, 67\% used Blocks repeatedly during the deployment, which is similar to the 74\% from the field study results. Finally, our survey responses indicated that participants strongly agreed that they enjoyed collaborating with others at the same time (\textit{M} = 6.26, \textit{SD} = 1.03). Our findings in this five-day deployment were in line with earlier studies on participation and end-user experience.

\section{Discussions and Design Implications}
Our research examined various design dimensions (Table~\ref{tab:DesignSpace}) and their effect on user experience. Our findings inform the design of future AR experiences. 

\subsection{Designing for Engagement and Participation}
Our findings suggest that more people will engage with a collaborative AR application in location-independent scenarios (Section~\ref{sec:fieldstudy_outcomeprocess}, Figure~\ref{fig:TimelineSession}), and that people prefer creating structures synchronously with colocated collaborators (Section~\ref{sec:labstudyresults_ux_fun}). Together, this suggests that a location-independent, colocated synchronous experience would bring both more usage and enjoyment.

\subsection{Promoting Creativity and Bonding}
In our lab and field studies, participants had fun, felt engaged, happier, more creative, and closer to their partner using Blocks (Section~\ref{sec:labstudyresults_ux_happy}, \ref{sec:labstudyresults_ux_fun}, \ref{sec:fieldstudy_ux}). For example, our lab study participants spent more time playing with Blocks than they were asked to (Section~\ref{sec:labstudyresults_ux_fun}). Also, 74\% of field study participants used Blocks more than once (Section~\ref{sec:fieldstudy_outcomeprocess}). Blocks showed how collaborative AR is a conduit for creativity, and enables people to create a wide range of AR structures and to get closer to their partners in the process.

\subsection{Supporting Creative Freedom through Onboarding}
Participants created a wider variety of structures and added more blocks when given an open-ended goal (Section~\ref{sec:labstudyresults_outcome}). The initial thematic goals were helpful as part of the onboarding experience, yet it gave people creative freedom. Furthermore, the personal sandbox allowed new users to explore Blocks more freely. Based on our findings, we encourage future designers to follow a three-staged onboarding process to support creative freedom: \textit{Stage 1}, enable a personal sandbox that provides a safe space for experimentation and exploration of features; \textit{Stage 2}, enable a limited shared space with specific goals; and \textit{Stage 3}, present a broader shared space with open-ended goals. 

\subsection{Facilitating Planning, Communication, and Awareness}
To collaborate effectively, participants often planned before creating with Blocks (Section~\ref{sec:labstudyresults_process_planning}). Participants in the lab study also left notes for future creators, clarifying their work or giving suggestions for improvements (Section~\ref{sec:labstudyresults_notes}). Designers should investigate tools to facilitate planning and communicating tips for future contributors.

Additionally, with Blocks' limited cursor representation, participants experienced difficulties locating each other in the 3D space while collaborating remotely (Section~\ref{sec:labstudyresults_process_communication}). Even when colocated, participants were still worried about getting in their partners' way (Section~\ref{sec:labstudyresults_process_movement}). We encourage future designers to enhance multi-user awareness by exploring techniques such as displaying out-of-sight cursors on the screen's edge, or integrating life-size avatars to simulate the colocated experience \cite{Orts-Escolano:2016}.

\subsection{Manipulating the Constraints of Reality}
We took a LEGO-like approach, which promoted physical movement and natural interactions working with physical blocks, e.g., participants stood on the chairs and laid on the ground to reach parts of the AR structure (Section~\ref{sec:labstudyresults_process_movement}). However, it also constrained the experience by physical reality, e.g., creation gets much harder once the structure is taller than the users. Therefore, depending on the goal, the experience could be designed to either simulate real-world affordances, or to manipulate the constraints of reality. To enable the latter, future work might explore creation tools such as editing groups of virtual objects in space and time \cite{spacetime}, or applying advanced graphics approaches to augment users' input, e.g., translating 2D line drawings to 3D structures around existing virtual/physical objects \cite{Krs:2017}.

\subsection{Leveraging Physical Environment}
According to the user experiences shared in our study results, participants took advantage of the physical environment while creating (Section~\ref{sec:fieldstudy_physicalenv}), although they felt limited by the tools provided by Blocks. Future designers should consider developing tools to support seamless interactions between the virtual and physical world. For example, models of objects in the scene could be captured and used for creation. Other properties such as color, text font, and texture could be used to make creation better integrated with the physical environment.

\subsection{Acknowledging the Role of Location}
In the location-dependent scenario, participants accessed Blocks fewer times than when they could use Blocks anywhere at any time (Figure~\ref{fig:TimelineSession}), even though \textit{``blocks adds a magical element of there being a persistent world taking place in parallel to what you see all day (P27).''} To encourage participation, we suggest future designers to separate consumption from creation, allowing users to consume content remotely, e.g., view AR content on a map, but limiting creation at only the specific location. Furthermore, users could be notified of changes to their creations to encourage them to return to the original location for further edits. To enable location-dependent scenarios in a naturalistic setting, future systems should explore using GPS for outdoor localization and Beacons for indoor positioning instead of using images as markers. With such capabilities, the world around us could be seamlessly interweaved with digital information, and be made more accessible \cite{Gleason:2016}.

In the location-independent scenario, participants reported feeling constrained by their surroundings when the AR structures were initially created in a larger space but later accessed in a smaller one (Section~\ref{sec:fieldstudy_prior}). Future systems could integrate mapping techniques, such as those in VR \cite{Sra:2018}, to make these experiences smoother.

\subsection{Improving Coordination and Motivation Support}
Participants used more blocks, on average, in the asynchronous scenario than in the synchronous one (Section~\ref{sec:labstudyresults_process_participation}). For example, one participant in the synchronous scenario felt that her creativity decreased because her partner was more skilled than her (Section~\ref{sec:labstudyresults_ux_happy}).
We believe that this synchronous presence of another participant might create ``process losses'', a term coined by Steiner and colleagues to describe actions, operations, or dynamics that prevent a group from reaching its full potential \cite{steiner2007group}. To address this, one could temporarily assign leadership roles to users \cite{Lasecki:2011}, or provide collaboration scaffolding through real-time prompts \cite{wang2017contrasting}. Additionally, one could apply group formation techniques to promote motivations between users and prevent social loafing \cite{latane1979many}.

\subsection{Considering Permissions and Content Moderation}
AR structures created using Blocks stay persistent for anyone to edit. Many participants in the lab study left notes for future creators not to remove or severely alter their creations, as this could mutilate their vision (Section~\ref{sec:labstudyresults_notes}). Participants who had expertise in information security and content moderation also raised concerns about potential vandalism of the shared AR world and user creations. Although we did not observe vandalism in our studies, future designers should consider employing more detailed permission control mechanisms (such as those in collaborative writing tools like Google Docs \cite{GoogleDocs}). This would enable, for example, families and friends to create a privately shared AR world. Additionally, as a system like this scales one should consider implementing content moderation techniques such as community reporting. 

\subsection{Exploring Expiry Settings}
Participants mentioned that the AR space got cluttered over time (Section~\ref{sec:fieldstudy_prior}) because structures persisted indefinitely. Designers should explore how to gradually disappear AR content that is not actively maintained (e.g., carelessly added blocks).

\section{Limitations and Future Work}
Our participants were always able to communicate, either by talking over the separator in the middle of the lab space, or via video conference in the case of the field deployments. In the future, designers should consider adding video and audio chatting functionality to the app to facilitate remote collaboration.

The Blocks app did not distinguish the contributions of different users. However, we observed participants claiming different parts of a structure when working together, or drawing their initials using virtual blocks to display authorship. In the future, we suggest exploring adding features to enable showing authorship, credit and provenance.

One limitation of our lab study design was that we featured varying tasks. Since the task type had a main effect (Section~\ref{sec:labstudyresults_outcome}), and our study design had to counterbalance for it, this caused variance in the results for collaboration modes. While this does not affect the validity of our results, it did make it more difficult to get quantitatively significant results. In the future, we might want to compare the collaboration more in more depth by using structurally equivalent tasks.

In this work, our study participants were predominantly young employees of a tech company, and over-represented early technology adopters. In the future, we plan to deploy this to a wider and more diverse population.

\section{Conclusion}
In this paper we introduced Blocks, a mobile AR application that enables collaborative and persistent experiences across multiple dimensions of space and time. Through a series of studies and deployments with over 160 participants, we found that participants created a wide variety of AR structures (Figure~\ref{fig:teaser}), had fun, felt engaged, and willingly spent more time working on their creations than required. In terms of preferences, participants enjoyed colocated synchronous collaboration the most. However, they were most active in the location-independent scenario where they participated in more sessions. To sum up, our findings suggest that creating location-independent, colocated synchronous experiences are more likely to lead to both enjoyable experiences and more frequent usage. Through our research, we identified a set of design implications for building future collaborative AR experiences. This paper articulates a vision of AR, where anyone can contribute to the virtual content overlaid on the world around us, making computing and information more accessible and embedded in the physical world.

\begin{acks}
Special thanks to Maarten Bos, Colin Eles, Mario Esparza, Erika Kerhwald, Patricia Mata, Isac Andreas M\"uller Sandvik, Brian Smith, Kyle Warner, Peicheng Yu, and Xu Wang for their help and support. Also, many thanks to all our study participants for their time, and the reviewers for their valuable feedback and suggestions.
\end{acks}

% Bibliography
\bibliographystyle{ACM-Reference-Format}
\bibliography{references}


\begin{thebibliography}{57}

%%% the Web address.

\ifx \showCODEN    \undefined \def \showCODEN     #1{\unskip}     \fi
\ifx \showDOI      \undefined \def \showDOI       #1{#1}\fi
\ifx \showISBNx    \undefined \def \showISBNx     #1{\unskip}     \fi
\ifx \showISBNxiii \undefined \def \showISBNxiii  #1{\unskip}     \fi
\ifx \showISSN     \undefined \def \showISSN      #1{\unskip}     \fi
\ifx \showLCCN     \undefined \def \showLCCN      #1{\unskip}     \fi
\ifx \shownote     \undefined \def \shownote      #1{#1}          \fi
\ifx \showarticletitle \undefined \def \showarticletitle #1{#1}   \fi
\ifx \showURL      \undefined \def \showURL       {\relax}        \fi
% invisible to TeX
\providecommand\bibfield[2]{#2}
\providecommand\bibinfo[2]{#2}
\providecommand\natexlab[1]{#1}
\providecommand\showeprint[2][]{arXiv:#2}

\bibitem[\protect\citeauthoryear{{Apple Inc.}}{{Apple Inc.}}{2019}]%
        {ARKit}
\bibfield{author}{\bibinfo{person}{{Apple Inc.}}}
  \bibinfo{year}{2019}\natexlab{}.
\newblock \bibinfo{title}{ARKit}.
\newblock
\newblock
\urldef\tempurl%
\url{https://developer.apple.com/arkit/}
\showURL{%
\tempurl}


\bibitem[\protect\citeauthoryear{{Archery Inc.}}{{Archery Inc.}}{2018}]%
        {arrow}
\bibfield{author}{\bibinfo{person}{{Archery Inc.}}}
  \bibinfo{year}{2018}\natexlab{}.
\newblock \bibinfo{title}{Arrow - Augmented Reality}.
\newblock
\newblock
\urldef\tempurl%
\url{https://itunes.apple.com/app/arrow-ar-texts-emojis/id1296755150}
\showURL{%
\tempurl}


\bibitem[\protect\citeauthoryear{Aron, Aron, and Smollan}{Aron
  et~al\mbox{.}}{1992}]%
        {aron1992inclusion}
\bibfield{author}{\bibinfo{person}{Arthur Aron}, \bibinfo{person}{Elaine~N
  Aron}, {and} \bibinfo{person}{Danny Smollan}.}
  \bibinfo{year}{1992}\natexlab{}.
\newblock \showarticletitle{Inclusion of other in the self scale and the
  structure of interpersonal closeness.}
\newblock \bibinfo{journal}{\emph{Journal of personality and social
  psychology}} \bibinfo{volume}{63}, \bibinfo{number}{4}
  (\bibinfo{year}{1992}), \bibinfo{pages}{596}.
\newblock


\bibitem[\protect\citeauthoryear{Benkler}{Benkler}{2006}]%
        {benkler2006wealth}
\bibfield{author}{\bibinfo{person}{Yochai Benkler}.}
  \bibinfo{year}{2006}\natexlab{}.
\newblock \bibinfo{booktitle}{\emph{The wealth of networks: How social
  production transforms markets and freedom}}.
\newblock \bibinfo{publisher}{Yale University Press}.
\newblock


\bibitem[\protect\citeauthoryear{Bhattacharyya, Nath, Jo, Jadhav, and
  Hammer}{Bhattacharyya et~al\mbox{.}}{2019}]%
        {brick}
\bibfield{author}{\bibinfo{person}{Po Bhattacharyya}, \bibinfo{person}{Radha
  Nath}, \bibinfo{person}{Yein Jo}, \bibinfo{person}{Ketki Jadhav}, {and}
  \bibinfo{person}{Jessica Hammer}.} \bibinfo{year}{2019}\natexlab{}.
\newblock \showarticletitle{Brick: Toward A Model for Designing Synchronous
  Colocated Augmented Reality Games}. In \bibinfo{booktitle}{\emph{Proceedings
  of the 2019 CHI Conference on Human Factors in Computing Systems}}
  \emph{(\bibinfo{series}{CHI '19})}. \bibinfo{publisher}{ACM},
  \bibinfo{address}{New York, NY, USA}, Article \bibinfo{articleno}{323},
  \bibinfo{numpages}{9}~pages.
\newblock
\showISBNx{978-1-4503-5970-2}
\urldef\tempurl%
\url{https://doi.org/10.1145/3290605.3300553}
\showDOI{\tempurl}


\bibitem[\protect\citeauthoryear{Billinghurst, Clark, and Lee}{Billinghurst
  et~al\mbox{.}}{2015}]%
        {Billinghurst:2015}
\bibfield{author}{\bibinfo{person}{Mark Billinghurst}, \bibinfo{person}{Adrian
  Clark}, {and} \bibinfo{person}{Gun Lee}.} \bibinfo{year}{2015}\natexlab{}.
\newblock \showarticletitle{A Survey of Augmented Reality}.
\newblock \bibinfo{journal}{\emph{Found. Trends Hum.-Comput. Interact.}}
  \bibinfo{volume}{8}, \bibinfo{number}{2-3} (\bibinfo{date}{March}
  \bibinfo{year}{2015}), \bibinfo{pages}{73--272}.
\newblock
\showISSN{1551-3955}
\urldef\tempurl%
\url{https://doi.org/10.1561/1100000049}
\showDOI{\tempurl}


\bibitem[\protect\citeauthoryear{Billinghurst, Poupyrev, Kato, and
  May}{Billinghurst et~al\mbox{.}}{2000}]%
        {billinghurst2000mixing}
\bibfield{author}{\bibinfo{person}{Mark Billinghurst}, \bibinfo{person}{Ivan
  Poupyrev}, \bibinfo{person}{Hirokazu Kato}, {and} \bibinfo{person}{Richard
  May}.} \bibinfo{year}{2000}\natexlab{}.
\newblock \showarticletitle{Mixing Realities in Shared Space: An Augmented
  Reality Interface for Collaborative Computing.}. In
  \bibinfo{booktitle}{\emph{IEEE International Conference on Multimedia and
  Expo (III)}}. Citeseer, \bibinfo{pages}{1641--1644}.
\newblock


\bibitem[\protect\citeauthoryear{Brockmann, Kr{\"u}ger, Stieglitz, and
  Bohlsen}{Brockmann et~al\mbox{.}}{2013}]%
        {brockmann2013framework}
\bibfield{author}{\bibinfo{person}{Tobias Brockmann}, \bibinfo{person}{Nina
  Kr{\"u}ger}, \bibinfo{person}{Stefan Stieglitz}, {and} \bibinfo{person}{Immo
  Bohlsen}.} \bibinfo{year}{2013}\natexlab{}.
\newblock \showarticletitle{A Framework for Collaborative Augmented Reality
  Applications}.
\newblock  (\bibinfo{year}{2013}).
\newblock


\bibitem[\protect\citeauthoryear{Cai, Wang, and Chiang}{Cai
  et~al\mbox{.}}{2014}]%
        {cai2014case}
\bibfield{author}{\bibinfo{person}{Su Cai}, \bibinfo{person}{Xu Wang}, {and}
  \bibinfo{person}{Feng-Kuang Chiang}.} \bibinfo{year}{2014}\natexlab{}.
\newblock \showarticletitle{A case study of Augmented Reality simulation system
  application in a chemistry course}.
\newblock \bibinfo{journal}{\emph{Computers in Human Behavior}}
  \bibinfo{volume}{37} (\bibinfo{year}{2014}), \bibinfo{pages}{31--40}.
\newblock


\bibitem[\protect\citeauthoryear{Damala, Cubaud, Bationo, Houlier, and
  Marchal}{Damala et~al\mbox{.}}{2008}]%
        {damala2008bridging}
\bibfield{author}{\bibinfo{person}{Areti Damala}, \bibinfo{person}{Pierre
  Cubaud}, \bibinfo{person}{Anne Bationo}, \bibinfo{person}{Pascal Houlier},
  {and} \bibinfo{person}{Isabelle Marchal}.} \bibinfo{year}{2008}\natexlab{}.
\newblock \showarticletitle{Bridging the gap between the digital and the
  physical: design and evaluation of a mobile augmented reality guide for the
  museum visit}. In \bibinfo{booktitle}{\emph{Proceedings of the 3rd
  international conference on Digital Interactive Media in Entertainment and
  Arts}}. ACM, \bibinfo{pages}{120--127}.
\newblock


\bibitem[\protect\citeauthoryear{{Experiments with Google}}{{Experiments with
  Google}}{2018}]%
        {GardenFriends}
\bibfield{author}{\bibinfo{person}{{Experiments with Google}}.}
  \bibinfo{year}{2018}\natexlab{}.
\newblock \bibinfo{title}{Garden Friends}.
\newblock
\newblock
\urldef\tempurl%
\url{https://experiments.withgoogle.com/garden-friends}
\showURL{%
\tempurl}


\bibitem[\protect\citeauthoryear{Gleason, Guo, Laput, Kitani, and
  Bigham}{Gleason et~al\mbox{.}}{2016}]%
        {Gleason:2016}
\bibfield{author}{\bibinfo{person}{Cole Gleason}, \bibinfo{person}{Anhong Guo},
  \bibinfo{person}{Gierad Laput}, \bibinfo{person}{Kris Kitani}, {and}
  \bibinfo{person}{Jeffrey~P. Bigham}.} \bibinfo{year}{2016}\natexlab{}.
\newblock \showarticletitle{VizMap: Accessible Visual Information Through
  Crowdsourced Map Reconstruction}. In \bibinfo{booktitle}{\emph{Proceedings of
  the 18th International ACM SIGACCESS Conference on Computers and
  Accessibility}} \emph{(\bibinfo{series}{ASSETS '16})}.
  \bibinfo{publisher}{ACM}, \bibinfo{address}{New York, NY, USA},
  \bibinfo{pages}{273--274}.
\newblock
\showISBNx{978-1-4503-4124-0}
\urldef\tempurl%
\url{https://doi.org/10.1145/2982142.2982200}
\showDOI{\tempurl}


\bibitem[\protect\citeauthoryear{Google}{Google}{2019}]%
        {GoogleDocs}
\bibfield{author}{\bibinfo{person}{Google}.} \bibinfo{year}{2019}\natexlab{}.
\newblock \bibinfo{title}{Google Docs}.
\newblock
\newblock
\urldef\tempurl%
\url{https://www.google.com/docs/about/}
\showURL{%
\tempurl}


\bibitem[\protect\citeauthoryear{{Google Developers}}{{Google
  Developers}}{2019}]%
        {ARCore}
\bibfield{author}{\bibinfo{person}{{Google Developers}}.}
  \bibinfo{year}{2019}\natexlab{}.
\newblock \bibinfo{title}{ARCore}.
\newblock
\newblock
\urldef\tempurl%
\url{https://developers.google.com/ar/}
\showURL{%
\tempurl}


\bibitem[\protect\citeauthoryear{Guest, MacQueen, and Namey}{Guest
  et~al\mbox{.}}{2011}]%
        {guest2011applied}
\bibfield{author}{\bibinfo{person}{Greg Guest}, \bibinfo{person}{Kathleen~M
  MacQueen}, {and} \bibinfo{person}{Emily~E Namey}.}
  \bibinfo{year}{2011}\natexlab{}.
\newblock \bibinfo{booktitle}{\emph{Applied thematic analysis}}.
\newblock \bibinfo{publisher}{sage}.
\newblock


\bibitem[\protect\citeauthoryear{Guo, Chen, Qi, White, Ghosh, Asakawa, and
  Bigham}{Guo et~al\mbox{.}}{2016}]%
        {Guo:2016}
\bibfield{author}{\bibinfo{person}{Anhong Guo},
  \bibinfo{person}{Xiang~`Anthony' Chen}, \bibinfo{person}{Haoran Qi},
  \bibinfo{person}{Samuel White}, \bibinfo{person}{Suman Ghosh},
  \bibinfo{person}{Chieko Asakawa}, {and} \bibinfo{person}{Jeffrey~P. Bigham}.}
  \bibinfo{year}{2016}\natexlab{}.
\newblock \showarticletitle{VizLens: A Robust and Interactive Screen Reader for
  Interfaces in the Real World}. In \bibinfo{booktitle}{\emph{Proceedings of
  the 29th Annual Symposium on User Interface Software and Technology}}
  \emph{(\bibinfo{series}{UIST '16})}. \bibinfo{publisher}{ACM},
  \bibinfo{address}{New York, NY, USA}, \bibinfo{pages}{651--664}.
\newblock
\showISBNx{978-1-4503-4189-9}
\urldef\tempurl%
\url{https://doi.org/10.1145/2984511.2984518}
\showDOI{\tempurl}


\bibitem[\protect\citeauthoryear{Guo, Kim, Chen, Yeh, Hudson, Mankoff, and
  Bigham}{Guo et~al\mbox{.}}{2017}]%
        {Guo:2017}
\bibfield{author}{\bibinfo{person}{Anhong Guo}, \bibinfo{person}{Jeeeun Kim},
  \bibinfo{person}{Xiang~`Anthony' Chen}, \bibinfo{person}{Tom Yeh},
  \bibinfo{person}{Scott~E. Hudson}, \bibinfo{person}{Jennifer Mankoff}, {and}
  \bibinfo{person}{Jeffrey~P. Bigham}.} \bibinfo{year}{2017}\natexlab{}.
\newblock \showarticletitle{Facade: Auto-generating Tactile Interfaces to
  Appliances}. In \bibinfo{booktitle}{\emph{Proceedings of the 2017 CHI
  Conference on Human Factors in Computing Systems}}
  \emph{(\bibinfo{series}{CHI '17})}. \bibinfo{publisher}{ACM},
  \bibinfo{address}{New York, NY, USA}, \bibinfo{pages}{5826--5838}.
\newblock
\showISBNx{978-1-4503-4655-9}
\urldef\tempurl%
\url{https://doi.org/10.1145/3025453.3025845}
\showDOI{\tempurl}


\bibitem[\protect\citeauthoryear{Ha, Woo, Lee, Lee, Ryu, Choi, and Lee}{Ha
  et~al\mbox{.}}{2010}]%
        {ha2010artalet}
\bibfield{author}{\bibinfo{person}{Taejin Ha}, \bibinfo{person}{Woontack Woo},
  \bibinfo{person}{Youngho Lee}, \bibinfo{person}{Junhun Lee},
  \bibinfo{person}{Jeha Ryu}, \bibinfo{person}{Hankyun Choi}, {and}
  \bibinfo{person}{Kwanheng Lee}.} \bibinfo{year}{2010}\natexlab{}.
\newblock \showarticletitle{ARtalet: tangible user interface based immersive
  augmented reality authoring tool for Digilog book}. In
  \bibinfo{booktitle}{\emph{Ubiquitous Virtual Reality (ISUVR), 2010
  International Symposium on}}. IEEE, \bibinfo{pages}{40--43}.
\newblock


\bibitem[\protect\citeauthoryear{Henrysson, Billinghurst, and Ollila}{Henrysson
  et~al\mbox{.}}{2005}]%
        {henrysson2005face}
\bibfield{author}{\bibinfo{person}{Anders Henrysson}, \bibinfo{person}{Mark
  Billinghurst}, {and} \bibinfo{person}{Mark Ollila}.}
  \bibinfo{year}{2005}\natexlab{}.
\newblock \showarticletitle{Face to face collaborative AR on mobile phones}. In
  \bibinfo{booktitle}{\emph{Proceedings of the 4th IEEE/ACM international
  symposium on mixed and augmented reality}}. IEEE Computer Society,
  \bibinfo{pages}{80--89}.
\newblock


\bibitem[\protect\citeauthoryear{{Inter IKEA Systems}}{{Inter IKEA
  Systems}}{2017}]%
        {IKEAPlace}
\bibfield{author}{\bibinfo{person}{{Inter IKEA Systems}}.}
  \bibinfo{year}{2017}\natexlab{}.
\newblock \bibinfo{title}{IKEA Place}.
\newblock
\newblock
\urldef\tempurl%
\url{https://highlights.ikea.com/2017/ikea-place/}
\showURL{%
\tempurl}


\bibitem[\protect\citeauthoryear{Jenkins}{Jenkins}{2006}]%
        {jenkins2006convergence}
\bibfield{author}{\bibinfo{person}{Henry Jenkins}.}
  \bibinfo{year}{2006}\natexlab{}.
\newblock \bibinfo{booktitle}{\emph{Convergence culture: Where old and new
  media collide}}.
\newblock \bibinfo{publisher}{NYU press}.
\newblock


\bibitem[\protect\citeauthoryear{Kim, Billinghurst, Bruder, Duh, and Welch}{Kim
  et~al\mbox{.}}{2018}]%
        {kim2018revisiting}
\bibfield{author}{\bibinfo{person}{Kangsoo Kim}, \bibinfo{person}{Mark
  Billinghurst}, \bibinfo{person}{Gerd Bruder}, \bibinfo{person}{Henry
  Been-Lirn Duh}, {and} \bibinfo{person}{Gregory~F Welch}.}
  \bibinfo{year}{2018}\natexlab{}.
\newblock \showarticletitle{Revisiting Trends in Augmented Reality Research: A
  Review of the 2nd Decade of ISMAR (2008--2017)}.
\newblock \bibinfo{journal}{\emph{IEEE transactions on visualization and
  computer graphics}} \bibinfo{volume}{24}, \bibinfo{number}{11}
  (\bibinfo{year}{2018}), \bibinfo{pages}{2947--2962}.
\newblock


\bibitem[\protect\citeauthoryear{Kiyokawa, Billinghurst, Hayes, Gupta, Sannohe,
  and Kato}{Kiyokawa et~al\mbox{.}}{2002}]%
        {kiyokawa2002communication}
\bibfield{author}{\bibinfo{person}{Kiyoshi Kiyokawa}, \bibinfo{person}{Mark
  Billinghurst}, \bibinfo{person}{Sohan~E Hayes}, \bibinfo{person}{Anoop
  Gupta}, \bibinfo{person}{Yuki Sannohe}, {and} \bibinfo{person}{Hirokazu
  Kato}.} \bibinfo{year}{2002}\natexlab{}.
\newblock \showarticletitle{Communication behaviors of co-located users in
  collaborative AR interfaces}. In \bibinfo{booktitle}{\emph{Proceedings of the
  1st International Symposium on Mixed and Augmented Reality}}. IEEE Computer
  Society, \bibinfo{pages}{139}.
\newblock


\bibitem[\protect\citeauthoryear{Krs, Yumer, Carr, Benes, and M\v{e}ch}{Krs
  et~al\mbox{.}}{2017}]%
        {Krs:2017}
\bibfield{author}{\bibinfo{person}{Vojt\v{e}ch Krs}, \bibinfo{person}{Ersin
  Yumer}, \bibinfo{person}{Nathan Carr}, \bibinfo{person}{Bedrich Benes}, {and}
  \bibinfo{person}{Radom\'{\i}r M\v{e}ch}.} \bibinfo{year}{2017}\natexlab{}.
\newblock \showarticletitle{Skippy: Single View 3D Curve Interactive Modeling}.
\newblock \bibinfo{journal}{\emph{ACM Trans. Graph.}} \bibinfo{volume}{36},
  \bibinfo{number}{4}, Article \bibinfo{articleno}{128} (\bibinfo{date}{July}
  \bibinfo{year}{2017}), \bibinfo{numpages}{12}~pages.
\newblock
\showISSN{0730-0301}
\urldef\tempurl%
\url{https://doi.org/10.1145/3072959.3073603}
\showDOI{\tempurl}


\bibitem[\protect\citeauthoryear{Lasecki, Murray, White, Miller, and
  Bigham}{Lasecki et~al\mbox{.}}{2011}]%
        {Lasecki:2011}
\bibfield{author}{\bibinfo{person}{Walter~S. Lasecki}, \bibinfo{person}{Kyle~I.
  Murray}, \bibinfo{person}{Samuel White}, \bibinfo{person}{Robert~C. Miller},
  {and} \bibinfo{person}{Jeffrey~P. Bigham}.} \bibinfo{year}{2011}\natexlab{}.
\newblock \showarticletitle{Real-time Crowd Control of Existing Interfaces}. In
  \bibinfo{booktitle}{\emph{Proceedings of the 24th Annual ACM Symposium on
  User Interface Software and Technology}} \emph{(\bibinfo{series}{UIST '11})}.
  \bibinfo{publisher}{ACM}, \bibinfo{address}{New York, NY, USA},
  \bibinfo{pages}{23--32}.
\newblock
\showISBNx{978-1-4503-0716-1}
\urldef\tempurl%
\url{https://doi.org/10.1145/2047196.2047200}
\showDOI{\tempurl}


\bibitem[\protect\citeauthoryear{Latan{\'e}, Williams, and Harkins}{Latan{\'e}
  et~al\mbox{.}}{1979}]%
        {latane1979many}
\bibfield{author}{\bibinfo{person}{Bibb Latan{\'e}}, \bibinfo{person}{Kipling
  Williams}, {and} \bibinfo{person}{Stephen Harkins}.}
  \bibinfo{year}{1979}\natexlab{}.
\newblock \showarticletitle{Many hands make light the work: The causes and
  consequences of social loafing.}
\newblock \bibinfo{journal}{\emph{Journal of personality and social
  psychology}} \bibinfo{volume}{37}, \bibinfo{number}{6}
  (\bibinfo{year}{1979}), \bibinfo{pages}{822}.
\newblock


\bibitem[\protect\citeauthoryear{Lave, Wenger, and Wenger}{Lave
  et~al\mbox{.}}{1991}]%
        {lave1991situated}
\bibfield{author}{\bibinfo{person}{Jean Lave}, \bibinfo{person}{Etienne
  Wenger}, {and} \bibinfo{person}{Etienne Wenger}.}
  \bibinfo{year}{1991}\natexlab{}.
\newblock \bibinfo{booktitle}{\emph{Situated learning: Legitimate peripheral
  participation}}. Vol.~\bibinfo{volume}{521423740}.
\newblock \bibinfo{publisher}{Cambridge university press Cambridge}.
\newblock


\bibitem[\protect\citeauthoryear{Lee, Kim, and Billinghurst}{Lee
  et~al\mbox{.}}{2005}]%
        {Lee:2005}
\bibfield{author}{\bibinfo{person}{Gun~A. Lee}, \bibinfo{person}{Gerard~J.
  Kim}, {and} \bibinfo{person}{Mark Billinghurst}.}
  \bibinfo{year}{2005}\natexlab{}.
\newblock \showarticletitle{Immersive Authoring: What You eXperience Is What
  You Get (WYXIWYG)}.
\newblock \bibinfo{journal}{\emph{Commun. ACM}} \bibinfo{volume}{48},
  \bibinfo{number}{7} (\bibinfo{date}{July} \bibinfo{year}{2005}),
  \bibinfo{pages}{76--81}.
\newblock
\showISSN{0001-0782}
\urldef\tempurl%
\url{https://doi.org/10.1145/1070838.1070840}
\showDOI{\tempurl}


\bibitem[\protect\citeauthoryear{Lee, Nelles, Billinghurst, and Kim}{Lee
  et~al\mbox{.}}{2004}]%
        {Lee:2004}
\bibfield{author}{\bibinfo{person}{Gun~A. Lee}, \bibinfo{person}{Claudia
  Nelles}, \bibinfo{person}{Mark Billinghurst}, {and}
  \bibinfo{person}{Gerard~Jounghyun Kim}.} \bibinfo{year}{2004}\natexlab{}.
\newblock \showarticletitle{Immersive Authoring of Tangible Augmented Reality
  Applications}. In \bibinfo{booktitle}{\emph{Proceedings of the 3rd IEEE/ACM
  International Symposium on Mixed and Augmented Reality}}
  \emph{(\bibinfo{series}{ISMAR '04})}. \bibinfo{publisher}{IEEE Computer
  Society}, \bibinfo{address}{Washington, DC, USA}, \bibinfo{pages}{172--181}.
\newblock
\showISBNx{0-7695-2191-6}
\urldef\tempurl%
\url{https://doi.org/10.1109/ISMAR.2004.34}
\showDOI{\tempurl}


\bibitem[\protect\citeauthoryear{Lindlbauer and Wilson}{Lindlbauer and
  Wilson}{2018}]%
        {remixedreality}
\bibfield{author}{\bibinfo{person}{David Lindlbauer} {and}
  \bibinfo{person}{Andy~D. Wilson}.} \bibinfo{year}{2018}\natexlab{}.
\newblock \showarticletitle{Remixed Reality: Manipulating Space and Time in
  Augmented Reality}. In \bibinfo{booktitle}{\emph{Proceedings of the 2018 CHI
  Conference on Human Factors in Computing Systems}}
  \emph{(\bibinfo{series}{CHI '18})}. \bibinfo{publisher}{ACM},
  \bibinfo{address}{New York, NY, USA}, Article \bibinfo{articleno}{129},
  \bibinfo{numpages}{13}~pages.
\newblock
\showISBNx{978-1-4503-5620-6}
\urldef\tempurl%
\url{https://doi.org/10.1145/3173574.3173703}
\showDOI{\tempurl}


\bibitem[\protect\citeauthoryear{Mabrito}{Mabrito}{2006}]%
        {mabrito2006study}
\bibfield{author}{\bibinfo{person}{Mark Mabrito}.}
  \bibinfo{year}{2006}\natexlab{}.
\newblock \showarticletitle{A study of synchronous versus asynchronous
  collaboration in an online business writing class}.
\newblock \bibinfo{journal}{\emph{The American Journal of Distance Education}}
  \bibinfo{volume}{20}, \bibinfo{number}{2} (\bibinfo{year}{2006}),
  \bibinfo{pages}{93--107}.
\newblock


\bibitem[\protect\citeauthoryear{{Microsoft}}{{Microsoft}}{2018}]%
        {Maquette}
\bibfield{author}{\bibinfo{person}{{Microsoft}}.}
  \bibinfo{year}{2018}\natexlab{}.
\newblock \bibinfo{title}{Microsoft Maquette}.
\newblock
\newblock
\urldef\tempurl%
\url{https://www.maquette.ms}
\showURL{%
\tempurl}


\bibitem[\protect\citeauthoryear{{Microsoft}}{{Microsoft}}{2019}]%
        {Minecraft}
\bibfield{author}{\bibinfo{person}{{Microsoft}}.}
  \bibinfo{year}{2019}\natexlab{}.
\newblock \bibinfo{title}{Minecraft}.
\newblock
\newblock
\urldef\tempurl%
\url{https://minecraft.net/}
\showURL{%
\tempurl}


\bibitem[\protect\citeauthoryear{M{\"u}ller, Kapadia, Frey, Klinger, Mann,
  Solenthaler, Sumner, and Gross}{M{\"u}ller et~al\mbox{.}}{2015}]%
        {muller2015statistical}
\bibfield{author}{\bibinfo{person}{Stephan M{\"u}ller},
  \bibinfo{person}{Mubbasir Kapadia}, \bibinfo{person}{Seth Frey},
  \bibinfo{person}{S Klinger}, \bibinfo{person}{Richard~P Mann},
  \bibinfo{person}{Barbara Solenthaler}, \bibinfo{person}{Robert~W Sumner},
  {and} \bibinfo{person}{Markus Gross}.} \bibinfo{year}{2015}\natexlab{}.
\newblock \showarticletitle{Statistical analysis of player behavior in
  Minecraft}. In \bibinfo{booktitle}{\emph{Proceedings of the 10th
  International Conference on the Foundations of Digital Games}}. Society for
  the Advancement of the Science of Digital Games.
\newblock


\bibitem[\protect\citeauthoryear{{Niantic, Inc.}}{{Niantic, Inc.}}{2016}]%
        {PokemonGo}
\bibfield{author}{\bibinfo{person}{{Niantic, Inc.}}}
  \bibinfo{year}{2016}\natexlab{}.
\newblock \bibinfo{title}{Pok\'emon Go}.
\newblock
\newblock
\urldef\tempurl%
\url{https://www.pokemongo.com/}
\showURL{%
\tempurl}


\bibitem[\protect\citeauthoryear{Olson, Wang, Olson, and Zhang}{Olson
  et~al\mbox{.}}{2017}]%
        {Olson:2017}
\bibfield{author}{\bibinfo{person}{Judith~S. Olson}, \bibinfo{person}{Dakuo
  Wang}, \bibinfo{person}{Gary~M. Olson}, {and} \bibinfo{person}{Jingwen
  Zhang}.} \bibinfo{year}{2017}\natexlab{}.
\newblock \showarticletitle{How People Write Together Now: Beginning the
  Investigation with Advanced Undergraduates in a Project Course}.
\newblock \bibinfo{journal}{\emph{ACM Trans. Comput.-Hum. Interact.}}
  \bibinfo{volume}{24}, \bibinfo{number}{1}, Article \bibinfo{articleno}{4}
  (\bibinfo{date}{March} \bibinfo{year}{2017}), \bibinfo{numpages}{40}~pages.
\newblock
\showISSN{1073-0516}
\urldef\tempurl%
\url{https://doi.org/10.1145/3038919}
\showDOI{\tempurl}


\bibitem[\protect\citeauthoryear{Orts-Escolano, Rhemann, Fanello, Chang,
  Kowdle, Degtyarev, Kim, Davidson, Khamis, Dou, Tankovich, Loop, Cai, Chou,
  Mennicken, Valentin, Pradeep, Wang, Kang, Kohli, Lutchyn, Keskin, and
  Izadi}{Orts-Escolano et~al\mbox{.}}{2016}]%
        {Orts-Escolano:2016}
\bibfield{author}{\bibinfo{person}{Sergio Orts-Escolano},
  \bibinfo{person}{Christoph Rhemann}, \bibinfo{person}{Sean Fanello},
  \bibinfo{person}{Wayne Chang}, \bibinfo{person}{Adarsh Kowdle},
  \bibinfo{person}{Yury Degtyarev}, \bibinfo{person}{David Kim},
  \bibinfo{person}{Philip~L. Davidson}, \bibinfo{person}{Sameh Khamis},
  \bibinfo{person}{Mingsong Dou}, \bibinfo{person}{Vladimir Tankovich},
  \bibinfo{person}{Charles Loop}, \bibinfo{person}{Qin Cai},
  \bibinfo{person}{Philip~A. Chou}, \bibinfo{person}{Sarah Mennicken},
  \bibinfo{person}{Julien Valentin}, \bibinfo{person}{Vivek Pradeep},
  \bibinfo{person}{Shenlong Wang}, \bibinfo{person}{Sing~Bing Kang},
  \bibinfo{person}{Pushmeet Kohli}, \bibinfo{person}{Yuliya Lutchyn},
  \bibinfo{person}{Cem Keskin}, {and} \bibinfo{person}{Shahram Izadi}.}
  \bibinfo{year}{2016}\natexlab{}.
\newblock \showarticletitle{Holoportation: Virtual 3D Teleportation in
  Real-time}. In \bibinfo{booktitle}{\emph{Proceedings of the 29th Annual
  Symposium on User Interface Software and Technology}}
  \emph{(\bibinfo{series}{UIST '16})}. \bibinfo{publisher}{ACM},
  \bibinfo{address}{New York, NY, USA}, \bibinfo{pages}{741--754}.
\newblock
\showISBNx{978-1-4503-4189-9}
\urldef\tempurl%
\url{https://doi.org/10.1145/2984511.2984517}
\showDOI{\tempurl}


\bibitem[\protect\citeauthoryear{Pan, Sinclair, and Mitchell}{Pan
  et~al\mbox{.}}{2018}]%
        {pan2018empowerment}
\bibfield{author}{\bibinfo{person}{Ye Pan}, \bibinfo{person}{David Sinclair},
  {and} \bibinfo{person}{Kenny Mitchell}.} \bibinfo{year}{2018}\natexlab{}.
\newblock \showarticletitle{Empowerment and embodiment for collaborative mixed
  reality systems}.
\newblock \bibinfo{journal}{\emph{Computer Animation and Virtual Worlds}}
  \bibinfo{volume}{29}, \bibinfo{number}{3-4} (\bibinfo{year}{2018}),
  \bibinfo{pages}{e1838}.
\newblock


\bibitem[\protect\citeauthoryear{Paucher and Turk}{Paucher and Turk}{2010}]%
        {paucher2010location}
\bibfield{author}{\bibinfo{person}{R{\'e}mi Paucher} {and}
  \bibinfo{person}{Matthew Turk}.} \bibinfo{year}{2010}\natexlab{}.
\newblock \showarticletitle{Location-based augmented reality on mobile phones}.
  In \bibinfo{booktitle}{\emph{Computer Vision and Pattern Recognition
  Workshops (CVPRW), 2010 IEEE Computer Society Conference on}}. IEEE,
  \bibinfo{pages}{9--16}.
\newblock


\bibitem[\protect\citeauthoryear{Rajagopal, Miller, Kumar, Luong, and
  Rowe}{Rajagopal et~al\mbox{.}}{2018}]%
        {Rajagopal:2018}
\bibfield{author}{\bibinfo{person}{Niranjini Rajagopal}, \bibinfo{person}{John
  Miller}, \bibinfo{person}{Krishna Kumar~Reghu Kumar}, \bibinfo{person}{Anh
  Luong}, {and} \bibinfo{person}{Anthony Rowe}.}
  \bibinfo{year}{2018}\natexlab{}.
\newblock \showarticletitle{Welcome to My World: Demystifying Multi-user AR
  with the Cloud: Demo Abstract}. In \bibinfo{booktitle}{\emph{Proceedings of
  the 17th ACM/IEEE International Conference on Information Processing in
  Sensor Networks}} \emph{(\bibinfo{series}{IPSN '18})}.
  \bibinfo{publisher}{IEEE Press}, \bibinfo{address}{Piscataway, NJ, USA},
  \bibinfo{pages}{146--147}.
\newblock
\showISBNx{978-1-5386-5298-5}
\urldef\tempurl%
\url{https://doi.org/10.1109/IPSN.2018.00036}
\showDOI{\tempurl}


\bibitem[\protect\citeauthoryear{Reitmayr and Schmalstieg}{Reitmayr and
  Schmalstieg}{2003}]%
        {Reitmayr:2003}
\bibfield{author}{\bibinfo{person}{Gerhard Reitmayr} {and}
  \bibinfo{person}{Dieter Schmalstieg}.} \bibinfo{year}{2003}\natexlab{}.
\newblock \showarticletitle{Location Based Applications for Mobile Augmented
  Reality}. In \bibinfo{booktitle}{\emph{Proceedings of the Fourth Australasian
  User Interface Conference on User Interfaces 2003 - Volume 18}}
  \emph{(\bibinfo{series}{AUIC '03})}. \bibinfo{publisher}{Australian Computer
  Society, Inc.}, \bibinfo{address}{Darlinghurst, Australia, Australia},
  \bibinfo{pages}{65--73}.
\newblock
\showISBNx{0-909925-96-8}
\urldef\tempurl%
\url{http://dl.acm.org/citation.cfm?id=820086.820103}
\showURL{%
\tempurl}


\bibitem[\protect\citeauthoryear{Resnick and Silverman}{Resnick and
  Silverman}{2005}]%
        {Resnick:2005}
\bibfield{author}{\bibinfo{person}{Mitchel Resnick} {and}
  \bibinfo{person}{Brian Silverman}.} \bibinfo{year}{2005}\natexlab{}.
\newblock \showarticletitle{Some Reflections on Designing Construction Kits for
  Kids}. In \bibinfo{booktitle}{\emph{Proceedings of the 2005 Conference on
  Interaction Design and Children}} \emph{(\bibinfo{series}{IDC '05})}.
  \bibinfo{publisher}{ACM}, \bibinfo{address}{New York, NY, USA},
  \bibinfo{pages}{117--122}.
\newblock
\showISBNx{1-59593-096-5}
\urldef\tempurl%
\url{https://doi.org/10.1145/1109540.1109556}
\showDOI{\tempurl}


\bibitem[\protect\citeauthoryear{Roth}{Roth}{1982}]%
        {roth1982ray}
\bibfield{author}{\bibinfo{person}{Scott~D Roth}.}
  \bibinfo{year}{1982}\natexlab{}.
\newblock \showarticletitle{Ray casting for modeling solids}.
\newblock \bibinfo{journal}{\emph{Computer graphics and image processing}}
  \bibinfo{volume}{18}, \bibinfo{number}{2} (\bibinfo{year}{1982}),
  \bibinfo{pages}{109--144}.
\newblock


\bibitem[\protect\citeauthoryear{Roussos, Johnson, Moher, Leigh, Vasilakis, and
  Barnes}{Roussos et~al\mbox{.}}{1999}]%
        {roussos1999learning}
\bibfield{author}{\bibinfo{person}{Maria Roussos}, \bibinfo{person}{Andrew
  Johnson}, \bibinfo{person}{Thomas Moher}, \bibinfo{person}{Jason Leigh},
  \bibinfo{person}{Christina Vasilakis}, {and} \bibinfo{person}{Craig Barnes}.}
  \bibinfo{year}{1999}\natexlab{}.
\newblock \showarticletitle{Learning and building together in an immersive
  virtual world}.
\newblock \bibinfo{journal}{\emph{Presence: Teleoperators \& Virtual
  Environments}} \bibinfo{volume}{8}, \bibinfo{number}{3}
  (\bibinfo{year}{1999}), \bibinfo{pages}{247--263}.
\newblock


\bibitem[\protect\citeauthoryear{{Snap, Inc.}}{{Snap, Inc.}}{2019}]%
        {Snapchat}
\bibfield{author}{\bibinfo{person}{{Snap, Inc.}}}
  \bibinfo{year}{2019}\natexlab{}.
\newblock \bibinfo{title}{Snapchat}.
\newblock
\newblock
\urldef\tempurl%
\url{https://whatis.snapchat.com/}
\showURL{%
\tempurl}


\bibitem[\protect\citeauthoryear{Sra, Mottelson, and Maes}{Sra
  et~al\mbox{.}}{2018}]%
        {Sra:2018}
\bibfield{author}{\bibinfo{person}{Misha Sra}, \bibinfo{person}{Aske
  Mottelson}, {and} \bibinfo{person}{Pattie Maes}.}
  \bibinfo{year}{2018}\natexlab{}.
\newblock \showarticletitle{Your Place and Mine: Designing a Shared VR
  Experience for Remotely Located Users}. In
  \bibinfo{booktitle}{\emph{Proceedings of the 2018 Designing Interactive
  Systems Conference}} \emph{(\bibinfo{series}{DIS '18})}.
  \bibinfo{publisher}{ACM}, \bibinfo{address}{New York, NY, USA},
  \bibinfo{pages}{85--97}.
\newblock
\showISBNx{978-1-4503-5198-0}
\urldef\tempurl%
\url{https://doi.org/10.1145/3196709.3196788}
\showDOI{\tempurl}


\bibitem[\protect\citeauthoryear{Starner, Mann, Rhodes, Levine, Healey, Kirsch,
  Picard, and Pentland}{Starner et~al\mbox{.}}{1997}]%
        {starner1997augmented}
\bibfield{author}{\bibinfo{person}{Thad Starner}, \bibinfo{person}{Steve Mann},
  \bibinfo{person}{Bradley Rhodes}, \bibinfo{person}{Jeffrey Levine},
  \bibinfo{person}{Jennifer Healey}, \bibinfo{person}{Dana Kirsch},
  \bibinfo{person}{Rosalind~W Picard}, {and} \bibinfo{person}{Alex Pentland}.}
  \bibinfo{year}{1997}\natexlab{}.
\newblock \showarticletitle{Augmented reality through wearable computing}.
\newblock \bibinfo{journal}{\emph{Presence: Teleoperators \& Virtual
  Environments}} \bibinfo{volume}{6}, \bibinfo{number}{4}
  (\bibinfo{year}{1997}), \bibinfo{pages}{386--398}.
\newblock


\bibitem[\protect\citeauthoryear{Steiner}{Steiner}{2007}]%
        {steiner2007group}
\bibfield{author}{\bibinfo{person}{Ivan~D Steiner}.}
  \bibinfo{year}{2007}\natexlab{}.
\newblock \showarticletitle{Group process and productivity (social
  psychological monograph)}.
\newblock  (\bibinfo{year}{2007}).
\newblock


\bibitem[\protect\citeauthoryear{{The LEGO Group}}{{The LEGO Group}}{2019}]%
        {lego}
\bibfield{author}{\bibinfo{person}{{The LEGO Group}}.}
  \bibinfo{year}{2019}\natexlab{}.
\newblock \bibinfo{title}{LEGO}.
\newblock
\newblock
\urldef\tempurl%
\url{https://www.lego.com/}
\showURL{%
\tempurl}


\bibitem[\protect\citeauthoryear{Tuite, Smith, and Studio}{Tuite
  et~al\mbox{.}}{2012}]%
        {tuite2012emergent}
\bibfield{author}{\bibinfo{person}{Kathleen Tuite}, \bibinfo{person}{Adam~M
  Smith}, {and} \bibinfo{person}{Expressive~Intelligence Studio}.}
  \bibinfo{year}{2012}\natexlab{}.
\newblock \showarticletitle{Emergent remix culture in an anonymous
  collaborative art system}. In \bibinfo{booktitle}{\emph{Proceedings of the
  8th AAAI Artificial Intelligence and Interactive Digital Entertainment
  Conference (Toronto, Canada, 2012)}}. \bibinfo{pages}{16--23}.
\newblock


\bibitem[\protect\citeauthoryear{Vaish, Gaikwad, Kovacs, Veit, Krishna,
  Arrieta~Ibarra, Simoiu, Wilber, Belongie, Goel, Davis, and Bernstein}{Vaish
  et~al\mbox{.}}{2017}]%
        {crowdresearch}
\bibfield{author}{\bibinfo{person}{Rajan Vaish}, \bibinfo{person}{Snehalkumar
  (Neil)~S. Gaikwad}, \bibinfo{person}{Geza Kovacs}, \bibinfo{person}{Andreas
  Veit}, \bibinfo{person}{Ranjay Krishna}, \bibinfo{person}{Imanol
  Arrieta~Ibarra}, \bibinfo{person}{Camelia Simoiu}, \bibinfo{person}{Michael
  Wilber}, \bibinfo{person}{Serge Belongie}, \bibinfo{person}{Sharad Goel},
  \bibinfo{person}{James Davis}, {and} \bibinfo{person}{Michael~S. Bernstein}.}
  \bibinfo{year}{2017}\natexlab{}.
\newblock \showarticletitle{Crowd Research: Open and Scalable University
  Laboratories}. In \bibinfo{booktitle}{\emph{Proceedings of the 30th Annual
  ACM Symposium on User Interface Software and Technology}}
  \emph{(\bibinfo{series}{UIST '17})}. \bibinfo{publisher}{ACM},
  \bibinfo{address}{New York, NY, USA}, \bibinfo{pages}{829--843}.
\newblock
\showISBNx{978-1-4503-4981-9}
\urldef\tempurl%
\url{https://doi.org/10.1145/3126594.3126648}
\showDOI{\tempurl}


\bibitem[\protect\citeauthoryear{Wang, Leach, and Lindeman}{Wang
  et~al\mbox{.}}{2013}]%
        {wang2013diy}
\bibfield{author}{\bibinfo{person}{Jia Wang}, \bibinfo{person}{Owen Leach},
  {and} \bibinfo{person}{Robert~W Lindeman}.} \bibinfo{year}{2013}\natexlab{}.
\newblock \showarticletitle{DIY World Builder: an immersive level-editing
  system}. In \bibinfo{booktitle}{\emph{3D User Interfaces (3DUI), 2013 IEEE
  Symposium on}}. IEEE, \bibinfo{pages}{195--196}.
\newblock


\bibitem[\protect\citeauthoryear{Wang, Wen, and Rose}{Wang
  et~al\mbox{.}}{2017}]%
        {wang2017contrasting}
\bibfield{author}{\bibinfo{person}{Xu Wang}, \bibinfo{person}{Miaomiao Wen},
  {and} \bibinfo{person}{Carolyn Rose}.} \bibinfo{year}{2017}\natexlab{}.
\newblock \showarticletitle{Contrasting explicit and implicit support for
  transactive exchange in team oriented project based learning}. In
  \bibinfo{booktitle}{\emph{International Conference on Computer Supported
  Collaborative Learning}} \emph{(\bibinfo{series}{CSCL '17})}.
  \bibinfo{publisher}{Philadelphia, PA: International Society of the Learning
  Sciences.}
\newblock


\bibitem[\protect\citeauthoryear{{wiARframe, Inc.}}{{wiARframe, Inc.}}{2018}]%
        {wiARframe}
\bibfield{author}{\bibinfo{person}{{wiARframe, Inc.}}}
  \bibinfo{year}{2018}\natexlab{}.
\newblock \bibinfo{title}{wiARframe}.
\newblock
\newblock
\urldef\tempurl%
\url{https://www.wiarframe.com}
\showURL{%
\tempurl}


\bibitem[\protect\citeauthoryear{Xia, Herscher, Perlin, and Wigdor}{Xia
  et~al\mbox{.}}{2018}]%
        {spacetime}
\bibfield{author}{\bibinfo{person}{Haijun Xia}, \bibinfo{person}{Sebastian
  Herscher}, \bibinfo{person}{Ken Perlin}, {and} \bibinfo{person}{Daniel
  Wigdor}.} \bibinfo{year}{2018}\natexlab{}.
\newblock \showarticletitle{Spacetime: Enabling Fluid Individual and
  Collaborative Editing in Virtual Reality}. In
  \bibinfo{booktitle}{\emph{Proceedings of the 31st Annual ACM Symposium on
  User Interface Software and Technology}} \emph{(\bibinfo{series}{UIST '18})}.
  \bibinfo{publisher}{ACM}, \bibinfo{address}{New York, NY, USA},
  \bibinfo{pages}{853--866}.
\newblock
\showISBNx{978-1-4503-5948-1}
\urldef\tempurl%
\url{https://doi.org/10.1145/3242587.3242597}
\showDOI{\tempurl}


\bibitem[\protect\citeauthoryear{{Zappar Ltd.}}{{Zappar Ltd.}}{2019}]%
        {ZapWorks}
\bibfield{author}{\bibinfo{person}{{Zappar Ltd.}}}
  \bibinfo{year}{2019}\natexlab{}.
\newblock \bibinfo{title}{ZapWorks}.
\newblock
\newblock
\urldef\tempurl%
\url{https://zap.works/}
\showURL{%
\tempurl}


\bibitem[\protect\citeauthoryear{Zhou, Duh, and Billinghurst}{Zhou
  et~al\mbox{.}}{2008}]%
        {zhou2008trends}
\bibfield{author}{\bibinfo{person}{Feng Zhou}, \bibinfo{person}{Henry Been-Lirn
  Duh}, {and} \bibinfo{person}{Mark Billinghurst}.}
  \bibinfo{year}{2008}\natexlab{}.
\newblock \showarticletitle{Trends in augmented reality tracking, interaction
  and display: A review of ten years of ISMAR}. In
  \bibinfo{booktitle}{\emph{Proceedings of the 7th IEEE/ACM International
  Symposium on Mixed and Augmented Reality}}. IEEE Computer Society,
  \bibinfo{pages}{193--202}.
\newblock


\end{thebibliography}

\end{document}